\documentclass[aps,prd,twocolumn,superscriptaddress,showpacs,preprintnumbers,amsmath,amssymb]{revtex4-2}

\usepackage{graphicx}
\usepackage{dcolumn}
\usepackage[format=plain,justification=centerlast,singlelinecheck=on]{caption}
\usepackage[format=plain,justification=justified,singlelinecheck=off]{subfig}
\usepackage[T1]{fontenc}

\graphicspath{{ps}}

\newcommand{\br}[1]{\ensuremath{\mathcal{B}\left(#1\right)}}

\begin{document}

\title{ \quad\\[1.0cm] Measurement of the branching fractions of the $B^+ \to \eta \ell^+ \nu_{\ell} $ and $B^+ \to \eta^{\prime} \ell^+ \nu_{\ell} $ decays with signal-side only reconstruction in the full $q^2$ range}

\noaffiliation
\affiliation{Department of Physics, University of the Basque Country UPV/EHU, 48080 Bilbao}
\affiliation{University of Bonn, 53115 Bonn}
\affiliation{Brookhaven National Laboratory, Upton, New York 11973}
\affiliation{Budker Institute of Nuclear Physics SB RAS, Novosibirsk 630090}
\affiliation{Faculty of Mathematics and Physics, Charles University, 121 16 Prague}
\affiliation{Chonnam National University, Gwangju 61186}
\affiliation{University of Cincinnati, Cincinnati, Ohio 45221}
\affiliation{Deutsches Elektronen--Synchrotron, 22607 Hamburg}
\affiliation{Duke University, Durham, North Carolina 27708}
\affiliation{Department of Physics, Fu Jen Catholic University, Taipei 24205}
\affiliation{Key Laboratory of Nuclear Physics and Ion-beam Application (MOE) and Institute of Modern Physics, Fudan University, Shanghai 200443}
\affiliation{Justus-Liebig-Universit\"at Gie\ss{}en, 35392 Gie\ss{}en}
\affiliation{II. Physikalisches Institut, Georg-August-Universit\"at G\"ottingen, 37073 G\"ottingen}
\affiliation{SOKENDAI (The Graduate University for Advanced Studies), Hayama 240-0193}
\affiliation{Gyeongsang National University, Jinju 52828}
\affiliation{Department of Physics and Institute of Natural Sciences, Hanyang University, Seoul 04763}
\affiliation{University of Hawaii, Honolulu, Hawaii 96822}
\affiliation{High Energy Accelerator Research Organization (KEK), Tsukuba 305-0801}
\affiliation{J-PARC Branch, KEK Theory Center, High Energy Accelerator Research Organization (KEK), Tsukuba 305-0801}
\affiliation{Higher School of Economics (HSE), Moscow 101000}
\affiliation{Forschungszentrum J\"{u}lich, 52425 J\"{u}lich}
\affiliation{IKERBASQUE, Basque Foundation for Science, 48013 Bilbao}
\affiliation{Indian Institute of Science Education and Research Mohali, SAS Nagar, 140306}
\affiliation{Indian Institute of Technology Bhubaneswar, Satya Nagar 751007}
\affiliation{Indian Institute of Technology Guwahati, Assam 781039}
\affiliation{Indian Institute of Technology Hyderabad, Telangana 502285}
\affiliation{Indian Institute of Technology Madras, Chennai 600036}
\affiliation{Indiana University, Bloomington, Indiana 47408}
\affiliation{Institute of High Energy Physics, Chinese Academy of Sciences, Beijing 100049}
\affiliation{Institute of High Energy Physics, Vienna 1050}
\affiliation{Institute for High Energy Physics, Protvino 142281}
\affiliation{INFN - Sezione di Napoli, 80126 Napoli}
\affiliation{INFN - Sezione di Torino, 10125 Torino}
\affiliation{J. Stefan Institute, 1000 Ljubljana}
\affiliation{Institut f\"ur Experimentelle Teilchenphysik, Karlsruher Institut f\"ur Technologie, 76131 Karlsruhe}
\affiliation{Kavli Institute for the Physics and Mathematics of the Universe (WPI), University of Tokyo, Kashiwa 277-8583}
\affiliation{Department of Physics, Faculty of Science, King Abdulaziz University, Jeddah 21589}
\affiliation{Kitasato University, Sagamihara 252-0373}
\affiliation{Korea Institute of Science and Technology Information, Daejeon 34141}
\affiliation{Korea University, Seoul 02841}
\affiliation{Kyungpook National University, Daegu 41566}
\affiliation{Universit\'{e} Paris-Saclay, CNRS/IN2P3, IJCLab, 91405 Orsay}
\affiliation{P.N. Lebedev Physical Institute of the Russian Academy of Sciences, Moscow 119991}
\affiliation{Faculty of Mathematics and Physics, University of Ljubljana, 1000 Ljubljana}
\affiliation{Ludwig Maximilians University, 80539 Munich}
\affiliation{Luther College, Decorah, Iowa 52101}
\affiliation{Malaviya National Institute of Technology Jaipur, Jaipur 302017}
\affiliation{Faculty of Chemistry and Chemical Engineering, University of Maribor, 2000 Maribor}
\affiliation{Max-Planck-Institut f\"ur Physik, 80805 M\"unchen}
\affiliation{School of Physics, University of Melbourne, Victoria 3010}
\affiliation{University of Mississippi, University, Mississippi 38677}
\affiliation{University of Miyazaki, Miyazaki 889-2192}
\affiliation{Moscow Physical Engineering Institute, Moscow 115409}
\affiliation{Graduate School of Science, Nagoya University, Nagoya 464-8602}
\affiliation{Kobayashi-Maskawa Institute, Nagoya University, Nagoya 464-8602}
\affiliation{Universit\`{a} di Napoli Federico II, 80126 Napoli}
\affiliation{Nara Women's University, Nara 630-8506}
\affiliation{National Central University, Chung-li 32054}
\affiliation{National United University, Miao Li 36003}
\affiliation{Department of Physics, National Taiwan University, Taipei 10617}
\affiliation{H. Niewodniczanski Institute of Nuclear Physics, Krakow 31-342}
\affiliation{Nippon Dental University, Niigata 951-8580}
\affiliation{Niigata University, Niigata 950-2181}
\affiliation{University of Nova Gorica, 5000 Nova Gorica}
\affiliation{Novosibirsk State University, Novosibirsk 630090}
\affiliation{Okinawa Institute of Science and Technology, Okinawa 904-0495}
\affiliation{Osaka City University, Osaka 558-8585}
\affiliation{Pacific Northwest National Laboratory, Richland, Washington 99352}
\affiliation{Panjab University, Chandigarh 160014}
\affiliation{Peking University, Beijing 100871}
\affiliation{University of Pittsburgh, Pittsburgh, Pennsylvania 15260}
\affiliation{Punjab Agricultural University, Ludhiana 141004}
\affiliation{Research Center for Nuclear Physics, Osaka University, Osaka 567-0047}
\affiliation{Meson Science Laboratory, Cluster for Pioneering Research, RIKEN, Saitama 351-0198}
\affiliation{Department of Modern Physics and State Key Laboratory of Particle Detection and Electronics, University of Science and Technology of China, Hefei 230026}
\affiliation{Seoul National University, Seoul 08826}
\affiliation{Showa Pharmaceutical University, Tokyo 194-8543}
\affiliation{Soochow University, Suzhou 215006}
\affiliation{Soongsil University, Seoul 06978}
\affiliation{Sungkyunkwan University, Suwon 16419}
\affiliation{School of Physics, University of Sydney, New South Wales 2006}
\affiliation{Department of Physics, Faculty of Science, University of Tabuk, Tabuk 71451}
\affiliation{Tata Institute of Fundamental Research, Mumbai 400005}
\affiliation{Department of Physics, Technische Universit\"at M\"unchen, 85748 Garching}
\affiliation{School of Physics and Astronomy, Tel Aviv University, Tel Aviv 69978}
\affiliation{Toho University, Funabashi 274-8510}
\affiliation{Department of Physics, Tohoku University, Sendai 980-8578}
\affiliation{Earthquake Research Institute, University of Tokyo, Tokyo 113-0032}
\affiliation{Department of Physics, University of Tokyo, Tokyo 113-0033}
\affiliation{Tokyo Institute of Technology, Tokyo 152-8550}
\affiliation{Tokyo Metropolitan University, Tokyo 192-0397}
\affiliation{Virginia Polytechnic Institute and State University, Blacksburg, Virginia 24061}
\affiliation{Wayne State University, Detroit, Michigan 48202}
\affiliation{Yamagata University, Yamagata 990-8560}
\affiliation{Yonsei University, Seoul 03722}
  \author{U.~Gebauer}\affiliation{II. Physikalisches Institut, Georg-August-Universit\"at G\"ottingen, 37073 G\"ottingen} 
  \author{C.~Bele\~{n}o}\affiliation{II. Physikalisches Institut, Georg-August-Universit\"at G\"ottingen, 37073 G\"ottingen} 
  \author{A.~Frey}\affiliation{II. Physikalisches Institut, Georg-August-Universit\"at G\"ottingen, 37073 G\"ottingen} 
  \author{I.~Adachi}\affiliation{High Energy Accelerator Research Organization (KEK), Tsukuba 305-0801}\affiliation{SOKENDAI (The Graduate University for Advanced Studies), Hayama 240-0193} 
  \author{K.~Adamczyk}\affiliation{H. Niewodniczanski Institute of Nuclear Physics, Krakow 31-342} 
  \author{H.~Aihara}\affiliation{Department of Physics, University of Tokyo, Tokyo 113-0033} 
  \author{S.~Al~Said}\affiliation{Department of Physics, Faculty of Science, University of Tabuk, Tabuk 71451}\affiliation{Department of Physics, Faculty of Science, King Abdulaziz University, Jeddah 21589} 
  \author{D.~M.~Asner}\affiliation{Brookhaven National Laboratory, Upton, New York 11973} 
  \author{H.~Atmacan}\affiliation{University of Cincinnati, Cincinnati, Ohio 45221} 
  \author{T.~Aushev}\affiliation{Higher School of Economics (HSE), Moscow 101000} 
% ...
\collaboration{The Belle Collaboration}

\begin{abstract}
The branching fractions of the decays $B^{+} \to \eta \ell^{+} \nu_{\ell}$ and $B^{+} \to \eta^{\prime} \ell^{+} \nu_{\ell}$ are measured, where $\ell$ is either an electron or a muon, using a data sample of $711\,{\rm fb}^{-1}$ containing $772 \times 10^6 B\bar{B}$ pairs collected at the $\Upsilon(4S)$ resonance with the Belle detector at the KEKB asymmetric-energy $e^+ e^-$ collider. To reduce the dependence of the result on the form factor model, the measurement is performed over the entire $q^2$ range. The resulting branching fractions are ${\cal B}(B^{+} \rightarrow \eta \ell^{+} \nu_{\ell}) = (2.83 \pm 0.55_{\rm (stat.)} \pm 0.34_{\rm (syst.)}) \times 10^{-5}$ and ${\cal B}(B^{+} \rightarrow \eta' \ell^{+} \nu_{\ell}) = (2.79 \pm 1.29_{\rm (stat.)} \pm 0.30_{\rm (syst.)}) \times 10^{-5}$.
\end{abstract}


\maketitle

\tighten

{\renewcommand{\thefootnote}{\fnsymbol{footnote}}}
\setcounter{footnote}{0}

\section{Introduction}
The transition $\overline{b} \to \overline{u} \ell^{+} \nu_{\ell}$, which spans over two generations of quarks, has been observed to be strongly suppressed. Understanding these decays is important to resolve the tension in the determination of the CKM matrix element $V_{\rm ub}$ by improving the model of inclusive $B \to X_{u} \ell^{+} \nu_{\ell}$ decays, and also the backgrounds in measurements of other decays. This paper describes the measurements of the branching fractions of the decays $B^{+} \to \eta \ell^{+} \nu_{\ell}$ and $B^{+} \to \eta^{\prime} \ell^{+} \nu_{\ell}$~\cite{CC}. Previous measurements of these decays typically restricted the measured range of the square of the momentum transfer ($q^{2} = (p_{B} - p_{\eta^{(\prime)}})^{2}$) which created a difficulty to quantify uncertainty in the modeling of the decay. This analysis reconstructs these decays without restrictions in $q^{2}$. Taking into account the hermeticity of the detector and the known initial state, only one of the two $B$ mesons produced in the decay of the $\Upsilon(4S)$ is reconstructed. This achieves the statistical power needed for studying such a suppressed set of processes. The modes analyzed in this paper have before been measured by BaBar~\cite{BabarEtaLoose,BabarEtaUntagged,BabarEtaSemilep}, CLEO~\cite{CLEOEtaPrime} and another Belle analysis~\cite{BelleEta} using hadronic tagging.

\section{The Belle detector and its data set}
The measurement presented here is based on the full data sample of $772 \times 10^6 B\overline{B}$ pairs collected  with the Belle detector at the KEKB asymmetric-energy $e^+e^-$ (3.5 on 8~GeV) collider~\cite{KEKB} operating at the $\Upsilon(4S)$ resonance. 

The Belle detector is a large-solid-angle magnetic spectrometer that consists of a silicon vertex detector (SVD), a 50-layer central drift chamber (CDC), an array of aerogel threshold Cherenkov counters (ACC),  
a barrel-like arrangement of time-of-flight scintillation counters (TOF), and an electromagnetic calorimeter comprised of CsI(Tl) crystals (ECL) located inside a superconducting solenoid coil that provides a 1.5~T
magnetic field.  An iron flux-return located outside of the coil is instrumented to detect $K_L^0$ mesons and to identify muons (KLM).  The detector is described in detail elsewhere~\cite{Belle}. Two inner detector configurations were used. A 2.0 cm radius beampipe and a 3-layer silicon vertex detector were used for the first sample of $152 \times 10^6 B\bar{B}$ pairs, while a 1.5 cm radius beampipe, a 4-layer silicon detector and a small-cell inner drift chamber were used to record the remaining $620 \times 10^6 B\bar{B}$ pairs~\cite{svd2}.\\

In this analysis several different sets of Monte Carlo (MC) simulated data have been used. Decays involving $b \to c$ transitions have been simulated with the equivalent of ten times the integrated luminosity acquired in data, while the transitions $e^+e^- \to q \bar{q},\; (q = u,d,s,c)$, denoted continuum, are simulated with six times the data luminosity. Additionally, a set containing one $B$ meson decaying via $\overline{b} \to \overline{u} \ell^{+} \nu_{\ell}$ with the other decaying via $b \to c$ is simulated with twenty times the integrated luminosity of data. This sample also contains the signal decay. The decays have been simulated using E\textsc{VT}G\textsc{EN}~\cite{EVTGEN} and PYTHIA~\cite{PYTHIA}, while the detector response was modeled with GEANT3~\cite{GEANT3}. Final-state radiation was added using PHOTOS~\cite{PHOTOS1}. The branching fractions for $B \to D^{(*)} \ell^{+} \nu_{\ell}$ decays, as well as exclusive and inclusive $\overline{b} \to \overline{u} \ell^{+} \nu_{\ell}$ decays have been updated to the most recent measurements~\cite{PDG}, except for the semileptonic decays to $\rho$ mesons, which have been set to the values measured by Ref.~\cite{rho0meas}. The form factors of the semileptonic decays to $D^{*}$ and $D^{**}$ have been updated according to the values reported in Ref.~\cite{HFLAV} and Ref.~\cite{LLSW}.

\section{Event selection and background suppression}\label{sec:selection}
All charged particle tracks and neutral clusters used in the analysis are required to satisfy basic quality criteria. Tracks need to originate from the interaction point (IP). All tracks for which the distance of closest approach to the IP in the longitudinal $|\mathrm{d}z|$ (perpendicular $|\mathrm{d}r|$) component with respect to the beam direction is greater than $2\; \mathrm{cm}$ ($0.5\; \mathrm{cm}$) are discarded. Tracks with a transverse momentum less than $275\;\mathrm{MeV}/c$ are checked for duplicates. A track is considered a duplicate if the three-momentum difference to another such track is less than $100\;\mathrm{MeV}/c$, and the angle between them is less than $15^{\circ}$ for equal charge or greater than $165^{\circ}$ for opposite charge tracks. For each such pair, only the track with the lesser value of $\left|5\times\mathrm{d}r\right|^{2}+\left|\mathrm{d}z\right|^{2}$ is kept. Photons are accepted within a polar angle, $\theta$, relative to the direction of the positron beam of $17^{\circ}$ to $150^{\circ}$. Due to variations in the distribution of beam-related backgrounds, different energy requirements are used depending on the polar angle region. In the central barrel region, from $32^{\circ}$ to $130^{\circ}$, photons are required to have an energy above $50 \;\mathrm{MeV}$. In the forward region, $\theta<32^{\circ}$, the requirement is $E_{\gamma}>100\;\mathrm{MeV}$,  while in the backward region, $\theta>130^{\circ}$, the requirement is $E_{\gamma}>150\;\mathrm{MeV}$, with the boundaries based on ECL geometry. As the total charge of the initial $e^+e^-$ system is zero, the sum of the charges of all reconstructed particles should be zero as well.  However, particles can be misreconstructed or completely escape detection. Therefore a requirement of $|\sum q_{\rm tracks}| < 3e$ is set.\\

A signal event is required to have only one lepton, which can be either an electron or a muon. Electrons must lie in the same acceptance region as photons with $17^{\circ} < \theta < 150^{\circ}$, while muons, for which KLM information is important, are accepted in the range $25^{\circ} < \theta < 145^{\circ}$. Both must have a center-of-mass (c.m.) momentum above $1.3 \; \mathrm{GeV}/c$, and electrons (muons) must have a lab frame momentum above $0.4 \;(0.8)\; \mathrm{GeV}/c$. Electrons are identified by using a likelihood function, which combines the shower shape in the ECL, the light yield in the ACC, the energy loss $dE/dx$ due to ionization in the CDC, the ratio of energy measured by the ECL to the momentum  of the track measured in the CDC, and quality of the matching of the CDC track to the ECL cluster position~\cite{eid}. The muon likelihood compares the CDC track with the associated KLM hits, using both the penetration length determined by the CDC and the matching quality of the KLM hits to the trajectory extrapolated out of the CDC~\cite{muid}. In the momentum region relevant to this analysis, charged leptons are identified with an efficiency of about $90\%$ and the probability to misidentify a pion as an electron (muon) is $0.25\%$ ($1.4\%$). For electron candidates, bremsstrahlung photons are recovered by searching photons in a $5^{\circ}$ cone around the electron. The closest photon is added to the lepton momentum unless it is used in the $\eta$ reconstruction.

The $\eta$ meson is reconstructed in two channels, $\eta \to \gamma \gamma$ and $\eta \to \pi^{+} \pi^{-} \pi^{0}$. A common background source for photons in the former is $\pi^{0}$ decays. A veto~\cite{pi0veto} against such photons is implemented by combining each candidate photon with all other photons in the event, and if the combined mass lies between $110 \;\mathrm{MeV}/c^2$ and $160 \;\mathrm{MeV}/c^2$, it is deemed to have come from a $\pi^{0}$ decay. Both photons in such a decay are discarded. All remaining photons are combined, where pairs with a mass between $510 \;\mathrm{MeV}/c^2$ and $580 \;\mathrm{MeV}/c^2$ are saved as $\eta$ candidates.\\

For the $\eta \to \pi^{+} \pi^{-} \pi^{0}$ channel, two pions of opposite charge are combined with one neutral pion built from two accepted photons. Charged pions are tested against the kaon hypothesis using a likelihood combining the energy loss $dE/dx$ from the CDC, the flight time measured by the TOF, and the ACC response, providing an efficiency of $86\%$ and a misidentification probability of $10\%$~\cite{pid}. The invariant mass of the combination is required to lie between $540 \;\mathrm{MeV}/c^2$ and $555 \;\mathrm{MeV}/c^2$. A vertex fit with $\chi^2/n.d.f. < 3$ is required for the $\eta$ candidate. The $\eta^{\prime}$ meson is reconstructed through the $\eta^\prime\to\pi^+\pi^-\eta$ decay mode. The $\eta$ candidates decay into two photons. The combined mass is required to be between $913 \;\mathrm{MeV}/c^2$ and $996 \;\mathrm{MeV}/c^2$, with the additional requirement that the mass difference $m_{\eta^{\prime}}-m_{\eta}$ must be between $400 \;\mathrm{MeV}/c^2$ and $420 \;\mathrm{MeV}/c^2$. Here too a vertex fit is required with $\chi^2/n.d.f. < 3$. The mass windows required for the $\eta$ and the $\eta^{\prime}$, as well as their mass difference, all correspond to a $\pm 3 \sigma$ window around the reconstructed value. Reconstruction of $\eta^{\prime}$ using $\eta \to \pi^{+} \pi^{-} \pi^{0}$ candidates proved to not be feasible due to increased combinatoric background.\\

\begin{figure}[htb]
\includegraphics[width=0.45\textwidth]{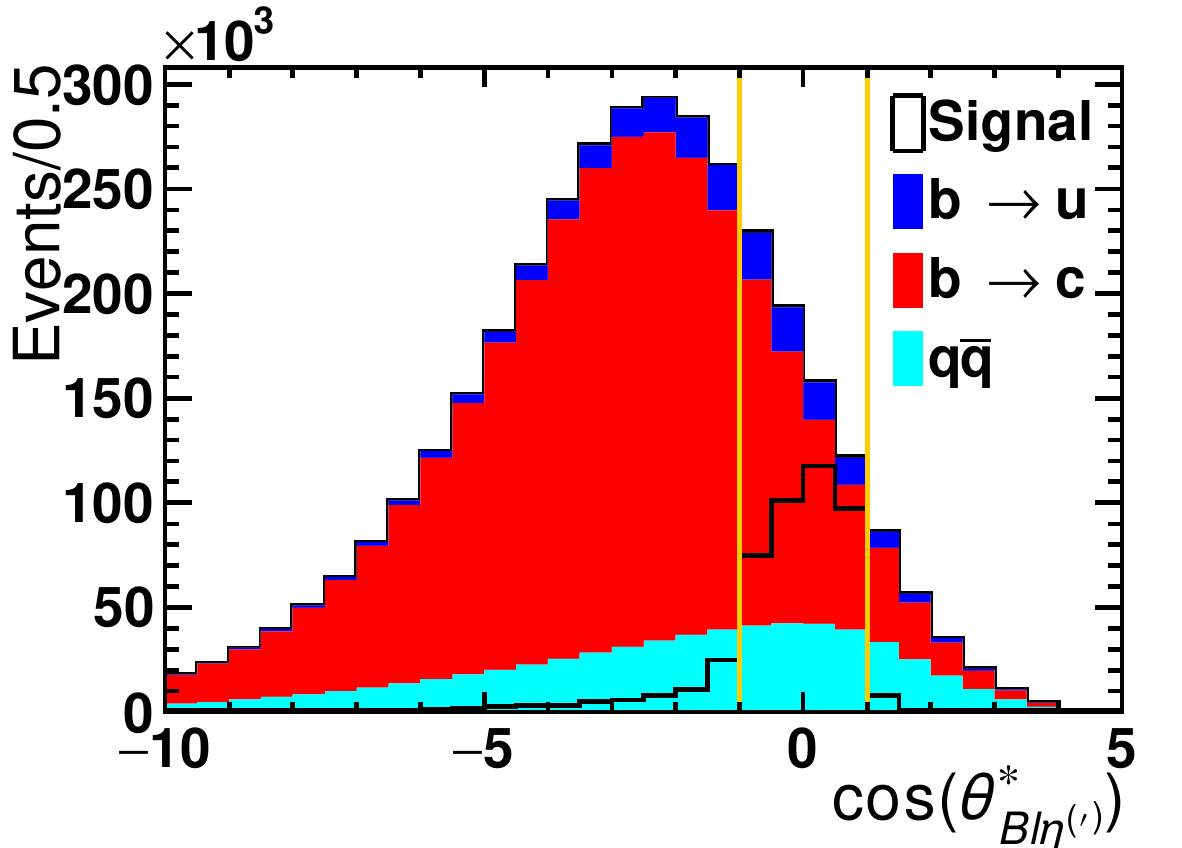}
\caption{Distribution of $\cos (\theta_{B\ell\eta^{(\prime)}}^{\star})$ for the $\eta\to\gamma\gamma$ channel, with all other requirements applied but before applying the BDT. Between the vertical yellow lines is the accepted region. The signal contribution is overlaid with an arbitrary scale factor. The distribution for the other channels looks very similar.}
\label{cosbyplot}
\end{figure}

Background is further suppressed using the angle between the $B$ meson and the combination of the lepton and $\eta^{(\prime)}$, defined as 
\begin{align}
\cos (\theta_{B\ell\eta^{(\prime)}}^{\star}) &= \frac{2E_B^{\star} E_{\ell\eta^{(\prime)}}^{\star} - m_B^2 c^4 -m_{\ell\eta^{(\prime)}}^2 c^4}{2 |\vec{p}^{\star}_B| |\vec{p}^{\star}_{\ell\eta^{(\prime)}}| c^2}, \label{eq:costheta}
\end{align}
where $E_B^{\star}$ and $|\vec{p}^{\star}_B|$ are the energy and momentum of the $B$ meson in the c.m.\ frame, $m_{\ell\eta^{(\prime)}}$ is the mass of the combined lepton-$\eta^{(\prime)}$ system, while $E_{\ell\eta^{(\prime)}}^{\star}$ and $|\vec{p}^{\star}_{\ell\eta^{(\prime)}}|$ are its energy and momentum. As the four-momentum of the $B$ meson can not be directly measured, the half of the c.m.\ energy $E^{\star}_B = \sqrt{s}/2$ and $|\vec{p}^{\star}_B| = \sqrt{s/4 - m_B^2 c^2}$ are used in Eq.~\ref{eq:costheta}. The distribution is shown in Fig.~\ref{cosbyplot}. All events must fulfil the requirements $|\cos (\theta_{B\ell\eta^{(\prime)}}^{\star})| < 1$, which ensures that they lie within the physical region. For remaining $B^+ \to \eta^{(\prime)} \ell^{+} \nu_{\ell}$ candidates, all charged decay products are fitted to a common vertex, and candidates for which the fit fails are discarded. Roughly half of the events contain more than one candidate in a single channel. Only the candidate with the lowest $\chi^2/n.d.f.$ from this fit is kept.\\

All final state particles of the signal decay chain have been measured except for the neutrino. Instead of a direct measurement, it is indirectly reconstructed with a technique~\cite{Aubert} using the known c.m.\ state of the event. Assuming all other particles in the event were detected, the difference between the sum of their 4-momenta and that of the initial state corresponds to the neutrino 4-momentum. This difference is called the missing momentum, $p_{\mathrm{miss}}$, and is defined as 
\begin{align}
p_{\mathrm{miss}} = p_{\Upsilon(4S)} - \left(\sum_{i}^{\mathrm{N}} E_i , \sum_{i}^{\mathrm{N}} \vec{p}_i \right),
\end{align}
where $p_{\Upsilon(4S)}$ is the momentum of the initial state of the $\Upsilon(4S)$, i.e.\ the sum of the two beam momenta, and energy and momenta of all $N$ particles remaining in the event are summed together. In this summation the requirements are loosened to $|dr| < 1.5 \;\mathrm{cm}$ and $|dz| < 10 \;\mathrm{cm}$. The invariant mass of the neutrino reconstructed in this way should be consistent with zero. Therefore, events with $|m_{\mathrm{miss}}^2| > 7\;\mathrm{GeV^2}/c^4$ are rejected. The missing mass is defined as:
\begin{align}
m_{\mathrm{miss}}^2 = |p_{\mathrm{miss}}|^2.
\end{align}
To compensate for the detector having better momentum resolution than the energy one, the neutrino energy in subsequent calculations is adjusted constraining $|E_{\nu_{\ell}}| = |\vec{p}_{\nu_{\ell}}|c$.\\

Using the inferred neutrino kinematics with the reconstructed $\ell$ and $\eta^{(\prime)}$ yields the $B^{\pm}$ candidate. Signal yield extraction uses the beam-constrained mass $M_{\mathrm{bc}} = \sqrt{ E_{\mathrm{beam}}^{\star 2}/c^4 - \vec{p}_B^{\star 2}/c^2 }$ and the energy difference $\Delta E = E_B^{\star} - E_{\mathrm{beam}}^{\star}$. Here $E_{\mathrm{beam}}^{\star}$ is the energy of one beam in the c.m.\ system, equivalent to half the c.m.\ energy, while $\vec{p}_{B}^{\star}$ and $E_{B}^{\star}$ are the three-momentum and energy of the combined $B$-daughter particles, including the inferred neutrino. At this point, candidates outside the fit region, $5.1 \;\mathrm{GeV}/c^2 < M_{\mathrm{bc}} < 5.3 \;\mathrm{GeV}/c^2 $ and $-1 \;\mathrm{GeV} < \Delta E < 1 \;\mathrm{GeV}$, are discarded.\\

\begin{table}[tbp]
\caption{Efficiencies of the event selection.}
\label{tabeff}
\begin{tabular*}{\linewidth}{@{\extracolsep{\fill}} @{\hspace{0.5cm}}lc@{\hspace{0.5cm}}}
\hline \hline
Channel & Efficiency (\%) \\ 
\hline
$\eta \to \gamma\gamma$    & 2.92\\ 
$\eta \to \pi^+\pi^-\pi^0$ & 2.03\\
$\eta^{\prime} \to \pi^+\pi^-\eta(\gamma\gamma)$ & 2.22\\ 
\hline \hline
\end{tabular*}
\end{table}

Continuum background is reduced by requiring the ratio of the second to zeroth Fox-Wolfram moment~\cite{SFW} to be less than $0.4$. Further background reduction uses boosted decision trees (BDTs)~\cite{TMVA}. For each channel, two such BDTs are trained, one to discriminate against background originating from $B\bar{B}$ events, and the other against continuum background. The training uses MC corresponding to the on-resonance integrated luminosity for the $b \to c$ and continuum background, and ten times the data integrated luminosity of the signal modes and the $\overline{b} \to \overline{u} \ell^{+} \nu_{\ell}$ background. These events are excluded from the analysis afterwards. The variables used by the BDT are: the number of particles with all general requirements applied except the distance to the IP; the number of charged particle tracks that fail the requirement on the distance to the IP; the number of $K^{\pm}$ candidates; $m_{\mathrm{miss}}^2$; the angles between the sum of all particles not assigned to the signal decay (representing the decay of the second $B$), and either the $\eta^{(\prime)}$ or the lepton candidate; the energy asymmetry of the two signal $\pi^{\pm}$ and the two $\gamma$ (where applicable) defined as $\mathcal{A}_{\eta} ~=~ ( E_{d1}-E_{d2} )/( E_{d1}+E_{d2} )$; and the difference between the squared momentum transfer, $q^2$, calculated with the inferred neutrino as $q^2 = (p_{\ell} + p_{\nu_{\ell}})^2$ and the method from Ref.~\cite{q2cleo}. The continuum classifier additionally uses the cosine of the angle between the thrust axes of the $\eta^{(\prime)}\ell^{+}$ system and the remaining event and 13 of the modified Fox-Wolfram moments~\cite{KSFW} found to be uncorrelated to $q^2$. The variable distributions are shown in Fig.~\ref{bdtvars1}, \ref{bdtvars2} and \ref{bdtvars6} in the appendix.\\

The selection of classifier input variables was restricted by the requirement to keep the entire $q^2$ range unbiased throughout the selection procedure. For each channel, the selections on the BDT output values are determined simultaneously by maximizing the figure of merit $N_{\rm Sig}/\sqrt{N_{\rm Sig}+N_{\rm Bkg}}$, where $N_{\rm Sig}(N_{\rm Bkg})$ is the number of signal (background) events in the remaining sample. The efficiencies of the event selection can be seen in Table~\ref{tabeff}. The agreement between data and MC was validated in sidebands. These sidebands consist of events outside the accepted $\eta$ mass range, or outside the range of the mass difference $m_{\eta^{\prime}}-m_{\eta}$ in case of the $\eta^{\prime}$. All other selection criteria including the BDT were unchanged. The signal region was only investigated after sufficient agreement in the description was verified.\\

\section{Signal determination}
The number of signal events in the remaining sample is determined with a two-dimensional binned maximum-likelihood fit~\cite{Barlow_finiteMC} in the variables $M_{\mathrm{bc}}$ and $\Delta E$ taking into account MC statistical uncertainties. Each $\eta^{(\prime)}$ channel is fitted individually, while no distinction between decays to electrons or muons is made in the fit. The fitted range is $5.1 \;\mathrm{GeV}/c^2 < M_{\mathrm{bc}} < 5.3 \;\mathrm{GeV}/c^2$ and $-1 \;\mathrm{GeV} < \Delta E < 1 \;\mathrm{GeV}$, divided into eight equal-sized bins in each variable. The four most signal-rich bins in the area $5.25 \;\mathrm{GeV} < M_{bc}$ and $|\Delta E| < 0.25\;\mathrm{GeV}$ further split into four bins each, giving a total of 76 bins for the fit. Pseudo-data generated from the MC have been used to validate the fit procedure, no bias in the results was observed.\\

\begin{figure*}[htpb]
\centering
\subfloat[$M_{\mathrm{bc}} (\eta\to\gamma\gamma)$]{
\includegraphics[width=0.3\textwidth ]{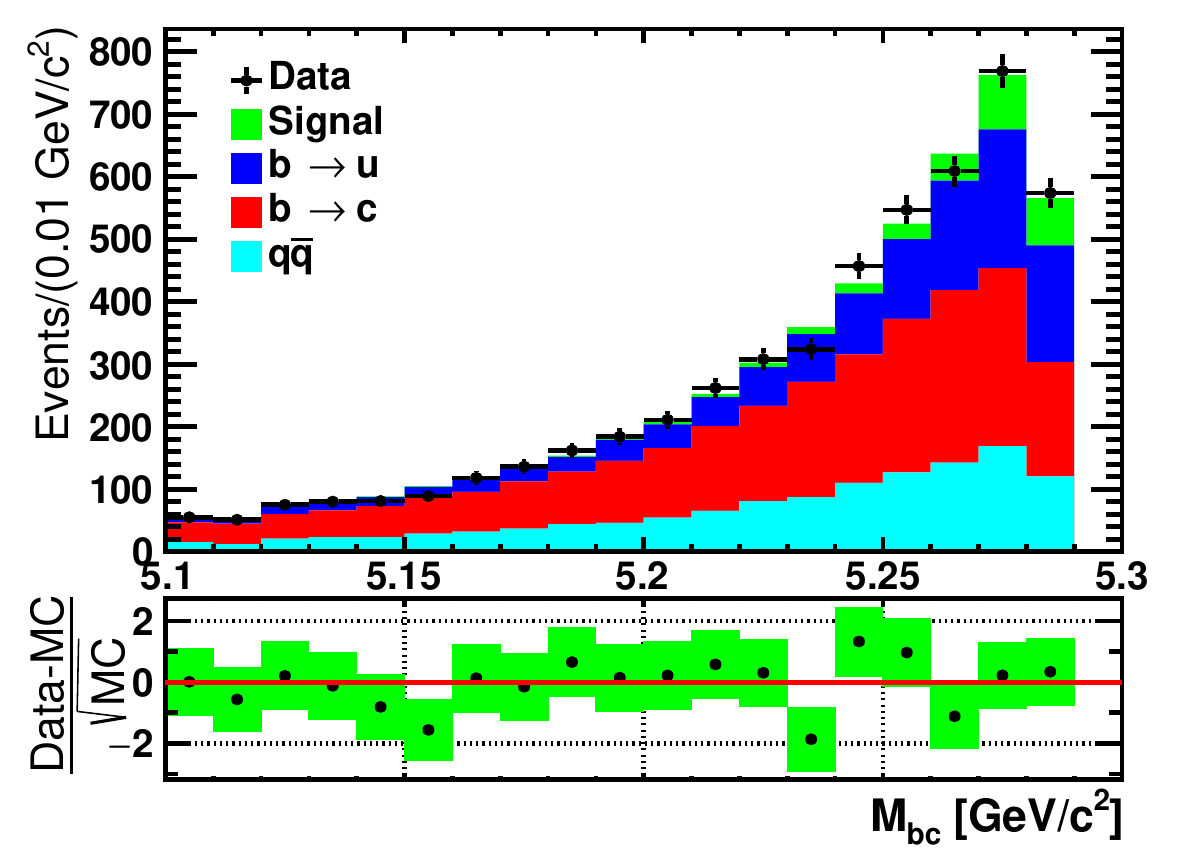}
}
\subfloat[$M_{\mathrm{bc}} (\eta\to\pi^+\pi^-\pi^0)$]{
\includegraphics[width=0.3\textwidth ]{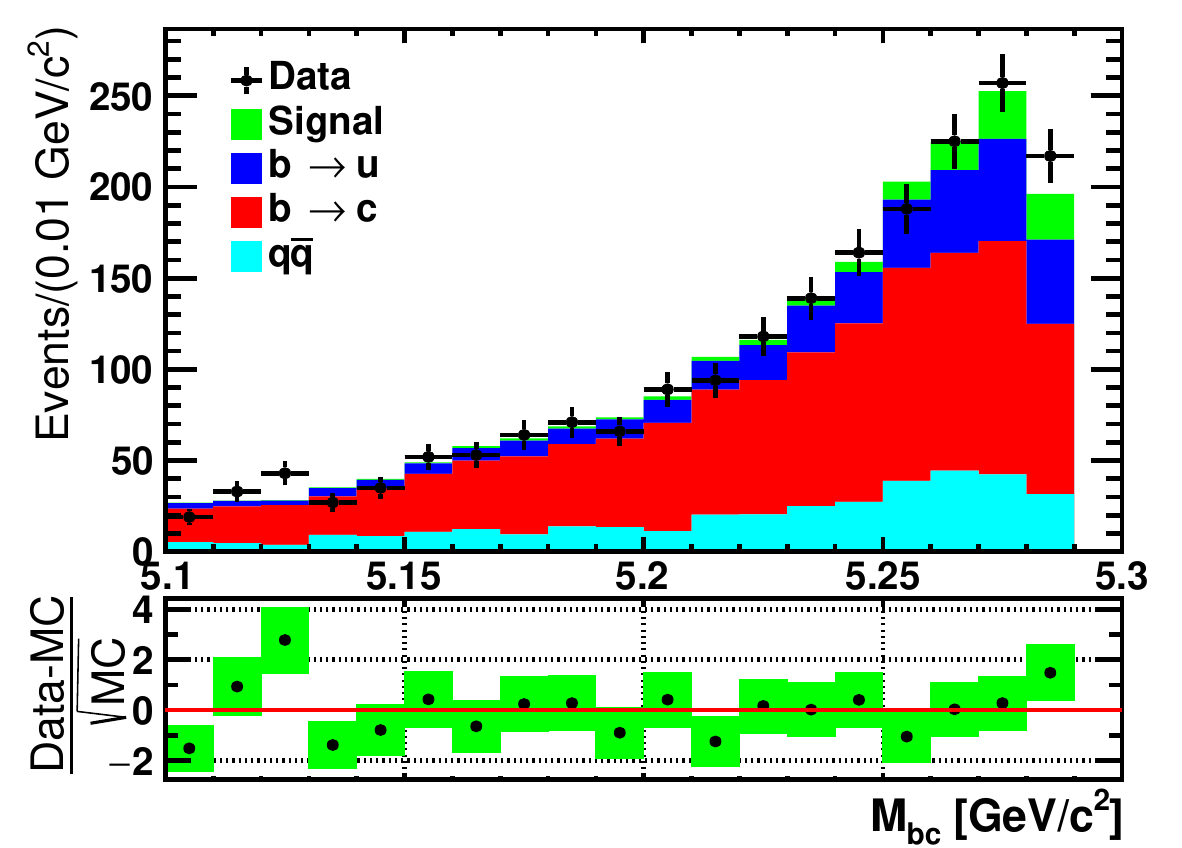}
}
\subfloat[$M_{\mathrm{bc}} (\eta^\prime\to\pi^+\pi^-\eta(\gamma\gamma))$]{
\includegraphics[width=0.3\textwidth ]{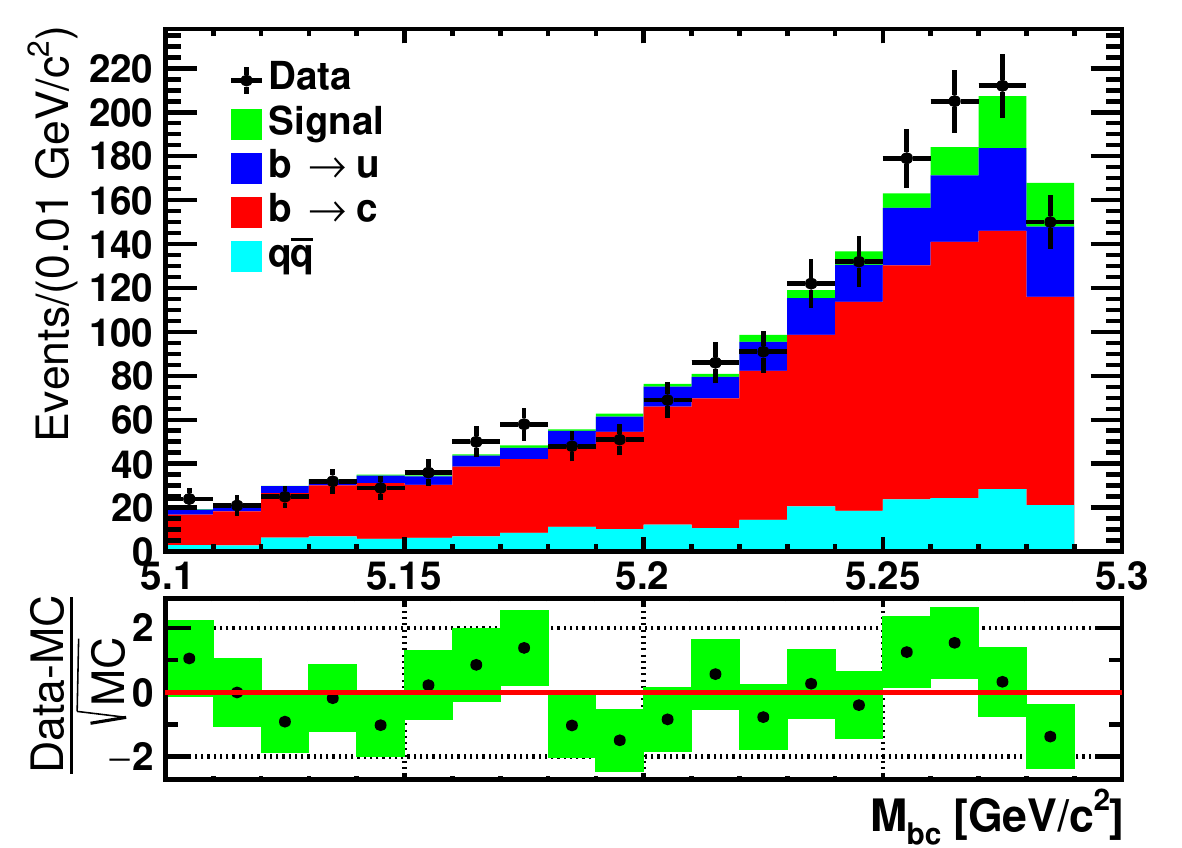}
}\\
\subfloat[$\Delta E (\eta\to\gamma\gamma)$]{
\includegraphics[width=0.3\textwidth ]{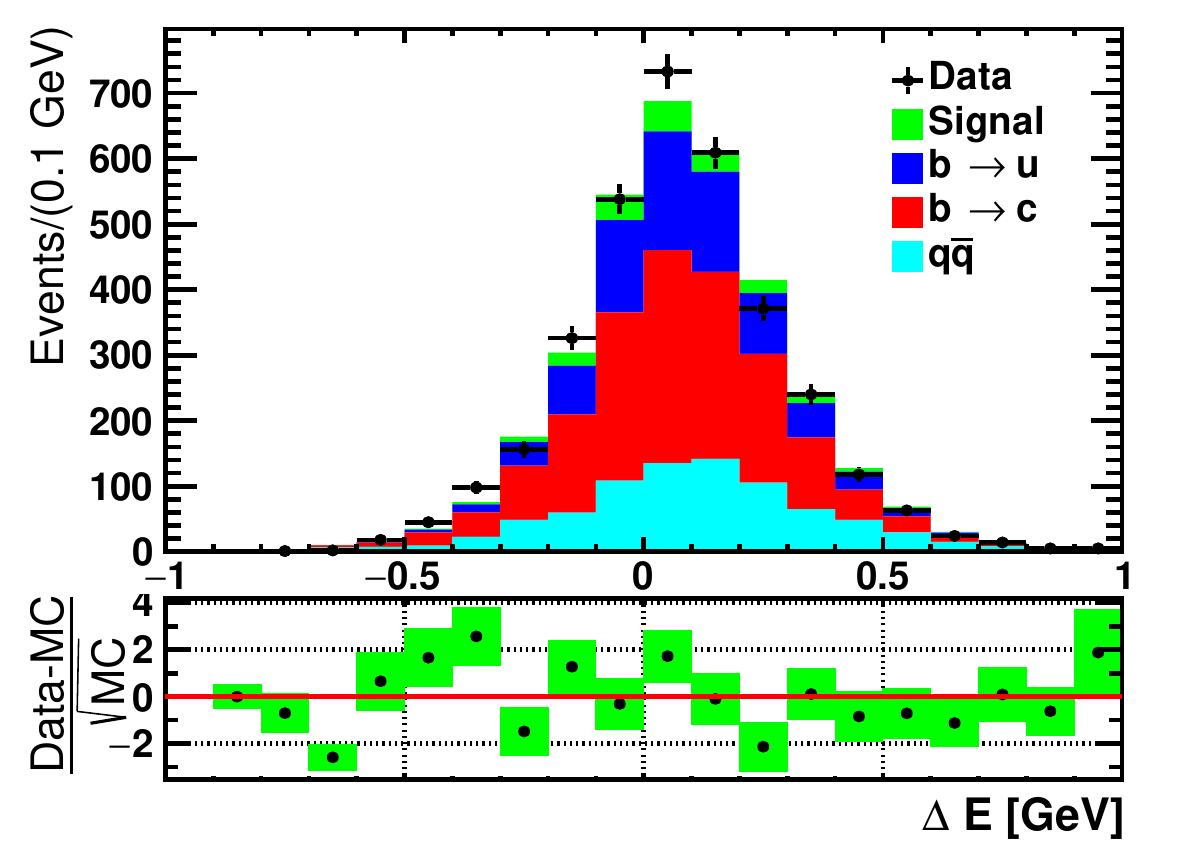}
}
\subfloat[$\Delta E (\eta\to\pi^+\pi^-\pi^0)$]{
\includegraphics[width=0.3\textwidth ]{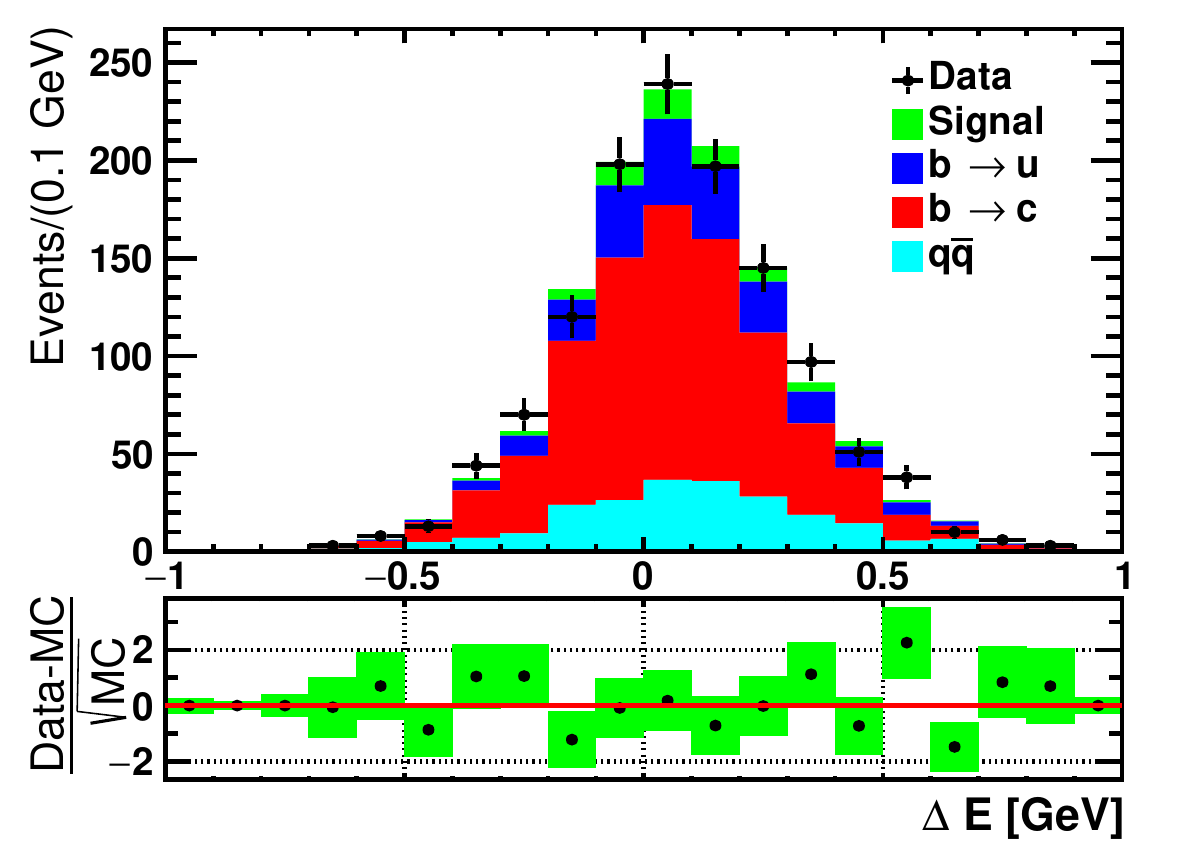}
}
\subfloat[$\Delta E (\eta^\prime\to\pi^+\pi^-\eta(\gamma\gamma))$]{
\includegraphics[width=0.3\textwidth ]{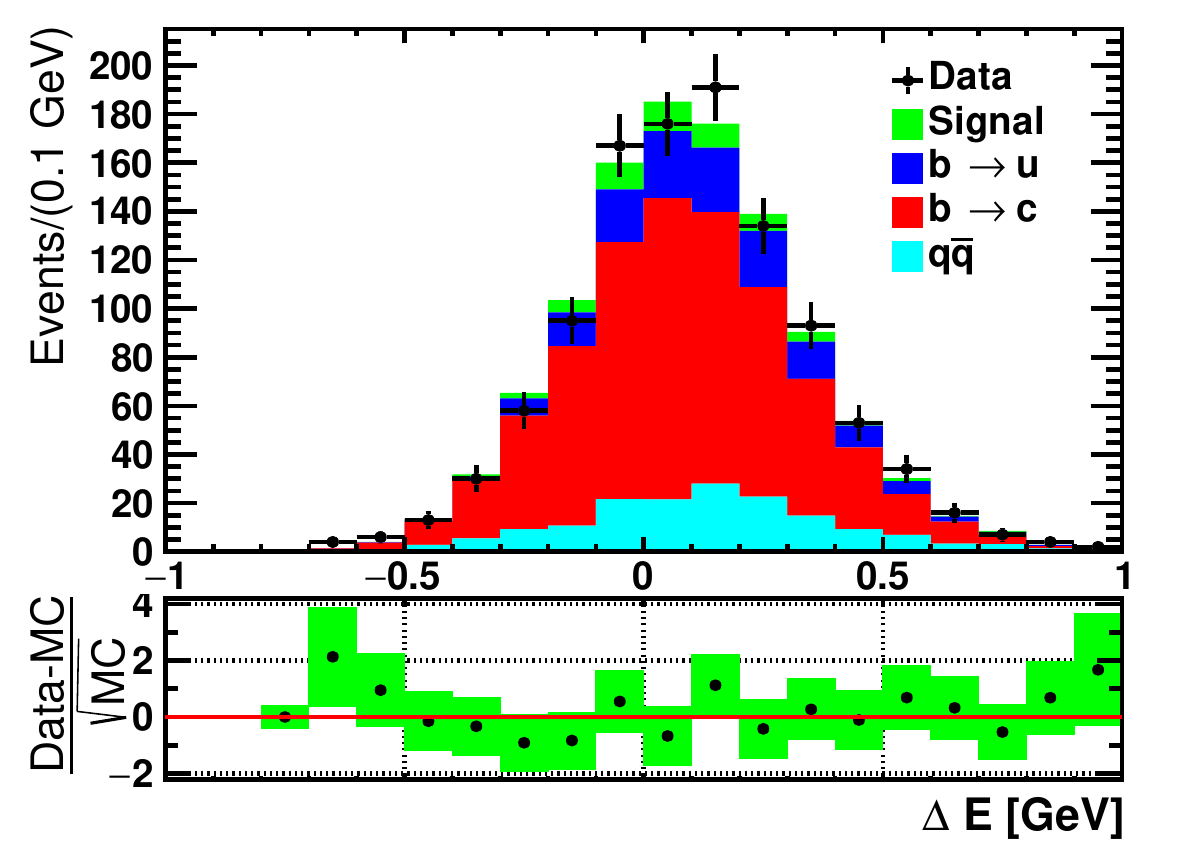}
}
\caption{Projections onto the two fit variables for all three channels used, with the contributions scaled to those obtained in the fit. The other variable is restricted to the signal-enriched region of $5.25 \;\mathrm{GeV}/c^2 < M_{\mathrm{bc}} < 5.3 \;\mathrm{GeV}/c^2 $ and $-0.25 \;\mathrm{GeV} < \Delta E < 0.25 \;\mathrm{GeV}$ respectively for visibility.}
\label{fitprojections}
\end{figure*}

The fit uses one signal and three background histogram templates. The first two background templates are the $B\bar{B}$ decays via $b \to c$ only and involving $b\to u$. The third is the continuum as defined in the BDT training. The contribution of the $\overline{b} \to \overline{u} \ell^{+} \nu_{\ell}$ background is fixed to the inclusive measurement from Ref.~\cite{HFLAV} while the other two backgrounds and the signal are determined by the fit. The number of events for each component and channel can be seen in Table~\ref{fityield} together with the efficiency $\epsilon$. The distributions resulting from the fit can be seen in Fig.~\ref{fitprojections}. With the fitted event yield, $N_{\mathrm{fit}}$ and the efficiency, $\epsilon$, the branching fraction is expressed as:
\begin{align}
&{\cal B}(B^+ \to \eta^{(\prime)} \ell^{+} \nu_{\ell}) = \nonumber\\ 
&\frac{N_{\mathrm{fit}}}{4 N_{B\bar{B}} {\cal B}(\Upsilon(4S) \to B^+B^-){\cal B}(\eta^{(\prime)} \to X)\epsilon}.\label{breq}
\end{align}
$\eta^{(\prime)} \to X$ denotes the decay of the $\eta^{(\prime)}$ into the respective final state. The sample contains $N_{B\bar{B}} = 772 \times 10^6$ pairs, the fraction of $B^+B^-$ among them is taken to be ${\cal B}(\Upsilon(4S) \to B^+B^-) = 0.513 \pm 0.006$~\cite{PDG} in this analysis. Both of these can decay to the signal mode. Together with the combination of decays to electrons and muons this gives a factor of four.\\

\begin{table}[htb]
\caption{Event yields, fit quality and selection efficiencies. For the fixed $\overline{b} \to \overline{u} \ell^{+} \nu_{\ell}$ component the Poissonian uncertainty of the yield is quoted.}
\label{fityield}
\begin{tabular*}{\linewidth}{@{\extracolsep{\fill}} @{\hspace{0.5cm}}lccc@{\hspace{0.5cm}}}
\hline \hline
Component & $\eta\to\gamma\gamma$ & $\eta\to\pi^+\pi^-\pi^0$ & $\eta'\to\pi^+\pi^-\eta$ \\
\hline
Signal      & 530 $\pm$ 116  & 196 $\pm$ 77   & 166 $\pm$ 76   \\ 
$\overline{b} \to \overline{u} \ell^{+} \nu_{\ell}$   & 2219 $\pm$ 47  & 674 $\pm$ 26   & 459 $\pm$ 22   \\ 
$b \to c$   & 4337 $\pm$ 233 & 2262 $\pm$ 147 & 2078 $\pm$ 150 \\ 
Continuum   & 2285 $\pm$ 221 & 692 $\pm$ 137  & 479 $\pm$ 129  \\
\hline
Data         & 9373           & 3828           & 3185            \\
\hline
$\chi^2/n.d.f.$ & $88.0/72$ & $86.3/72$ &  $64.4/72$         \\
\hline
Efficiency & 2.92\% & 2.03\% & 2.23\%\\
\hline \hline
\end{tabular*}
\end{table}

\section{Systematic uncertainties}
The sources of systematic uncertainty considered for this analysis fall into three categories, uncertainties related (i) to the detector performance, (ii) to the quality of the MC model and related input parameters, and (iii) to the fit procedure used. Unless otherwise stated, these are estimated by varying the relevant parameter by one standard deviation and repeating the full study. All individual components listed here are assumed to be uncorrelated. The values of the individual contributions are shown in Table~\ref{tab:syst:exp}.

\subsection{MC modeling and theory}
The most important background source for this analysis is other semileptonic decays of the type $B \to \ell^{+} \nu_{\ell} X_{Bkg}$. Their branching fractions are re-weighted to the current values taken from~\cite{PDG} and are varied within their uncertainty.\\

Semileptonic decay form factors are another important input of the MC model for which uncertainties are estimated. The form factors for the $b \to c$ decays $B \to D \ell^{+} \nu_{\ell}$, $B \to D^{*} \ell^{+} \nu_{\ell}$, $B \to D_1 \ell^{+} \nu_{\ell}$, $B \to D_0 \ell^{+} \nu_{\ell}$, $B \to D_{0}' \ell^{+} \nu_{\ell}$, and $B \to D_2 \ell^{+} \nu_{\ell}$ have been updated to the most recent values~\cite{HFLAV,LLSW} with the method described in Ref.~\cite{DFF} for both charged and neutral decaying $B$ mesons. The signal decays $B^{+} \to \eta \ell^{+} \nu_{\ell}$ and $B^{+} \to \eta^{\prime} \ell^{+} \nu_{\ell}$ are reweighted from the ISGW2 model~\cite{ISGW2} to the model taken from Ref.~\cite{Etareweight} with the form factors updated to Ref.~\cite{EtaFF}, using the BZ parametrisation and assuming uncorrelated parameters. The decay $B^{+} \to \omega \ell^{+} \nu_{\ell}$ is modeled according to Ref.~\cite{oldball} in the MC used and reweighted to Ref.~\cite{newball} for comparison. The shape of the inclusive component~\cite{inclshape} of the $\overline{b} \to \overline{u} \ell^{+} \nu_{\ell}$ transitions is also considered. The form factor uncertainties listed in Table~\ref{tab:syst:exp} are based on those reported in the publications they were obtained from. Despite having a slowly varying efficiency the $\eta \to \gamma \gamma$ mode appears to have the largest such uncertainty.\\

The effect of remaining background events containing $K_{L}^{0}$ is considered by varying the yield of such events up and down by $20\%$ when building the MC templates for the fit. Missing momentum indicating a neutrino can be faked by $K_{L}^{0}$. The continuum MC consists of two separate components, decays via a $c\bar{c}$ pair and those via a pair of the three lighter quarks. Effects of a mismodeled continuum are included by varying the ratio of the two components by $20\%$. The total measured number of $B\bar{B}$ pairs has an uncertainty of $1.4\%$ which is propagated through Eq.~\ref{breq}, as do the branching fractions of $\Upsilon(4S) \to B^+B^-$ and the subsequent signal decay chain taken from Ref.~\cite{PDG}. The MC statistics are assumed to have Poisson-distributed uncertainties due to their finite size.\\

\subsection{Detector performance}
Independent data samples have been used to validate the detector description and detection efficiency in the MC. A fully correlated uncertainty of $0.35\%$ per charged particle due to track recognition and $2\%$ per photon used in the signal reconstruction is assigned. For $\pi^0$ candidates a combined uncertainty of $2.5\%$ is assigned instead. Studies on the performance of the particle identification (PID) for both charged leptons and pions led to the use of angle and momentum-dependent correction factors with an associated uncertainty. The two pions are ordered by their energy with the first pion always being the higher-energy one. Additionally, the yield of background events with a lepton candidate not being an actual lepton originating directly from a $B$ meson decay is varied by $20\%$ to estimate the effect of incorrectly assigning the lepton source.\\

\begin{figure*}[htpb]
\centering
\subfloat[$\eta\to\gamma\gamma$]{
\includegraphics[width=0.3\textwidth ]{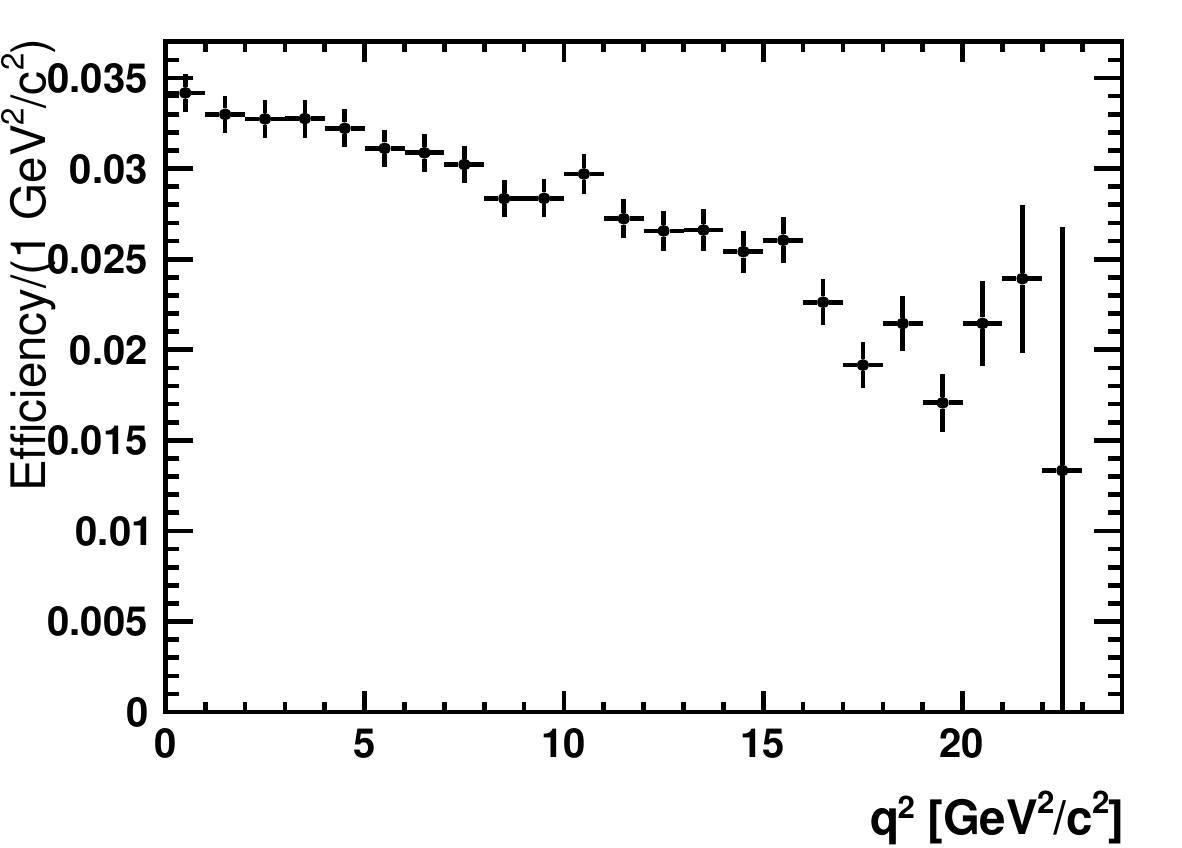}
}
\subfloat[$\eta\to\pi^+\pi^-\pi^0$]{
\includegraphics[width=0.3\textwidth ]{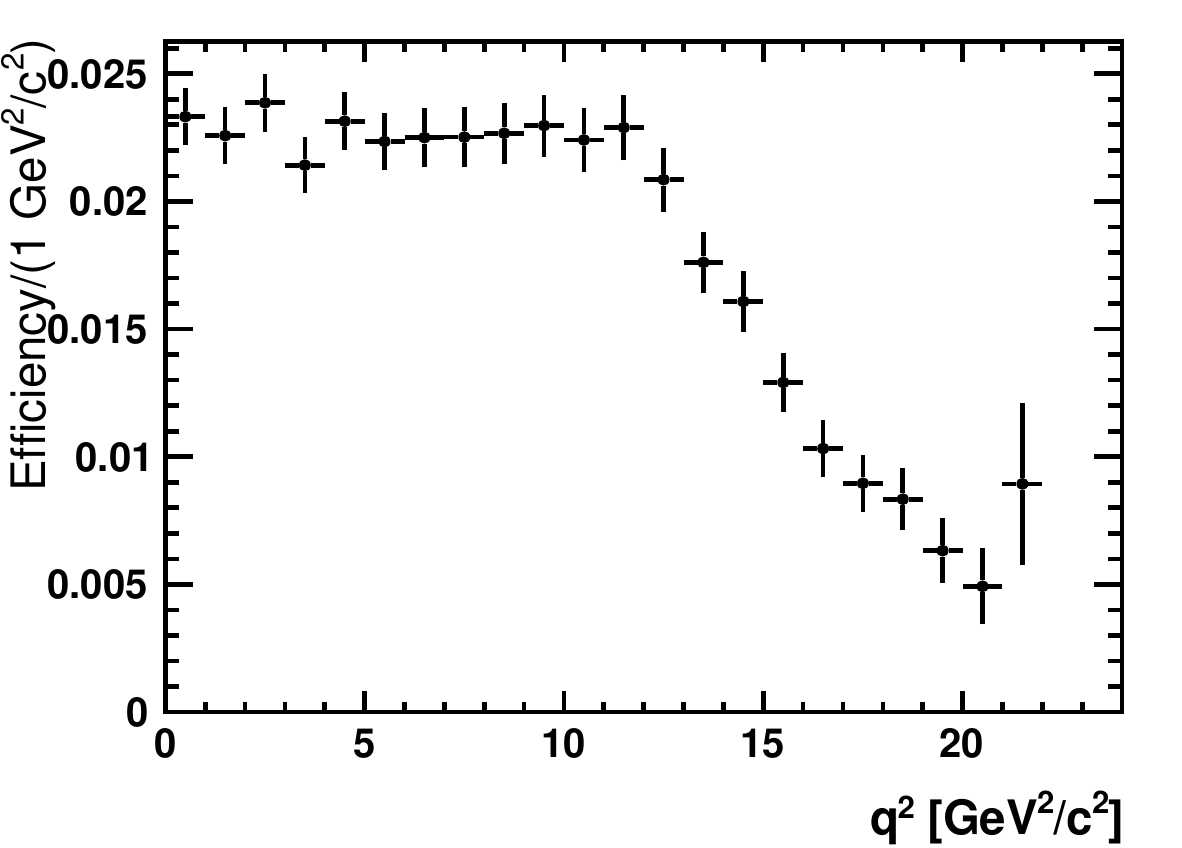}
}
\subfloat[$\eta^\prime\to\pi^+\pi^-\eta(\gamma\gamma)$]{
\includegraphics[width=0.3\textwidth ]{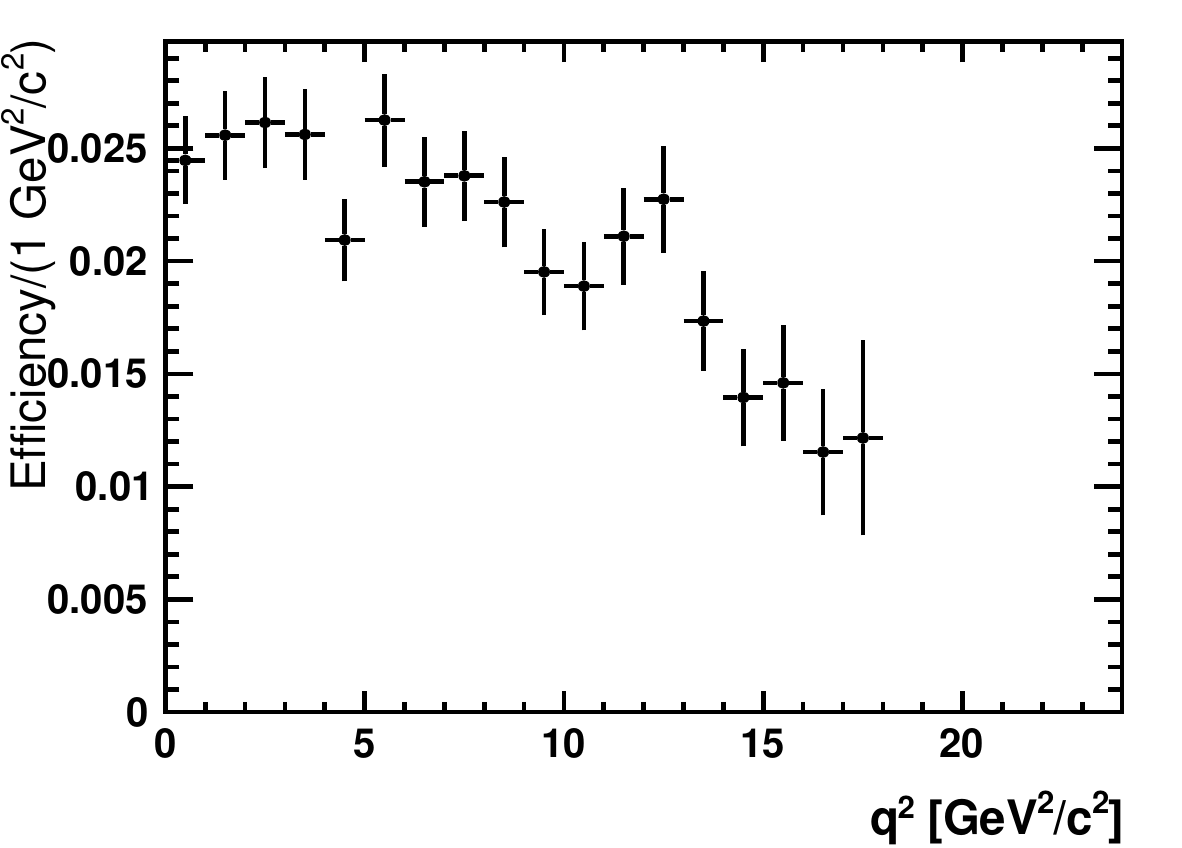}
}\\
\caption{Efficiency of the entire reconstruction chain including the BDT as a function of $q^2$}
\label{q2eff}
\end{figure*}

The dependence of the reconstruction efficiency on the value of the momentum transfer $q^2$ for an event is shown in Fig.~\ref{q2eff}. The $\eta\to\gamma\gamma$ channel only shows a weak dependence on $q^2$, while the other two channels show a decrease at large $q^2$ values which can be traced to pion detection efficiency.\\

\subsection{Fit validation}
The much more common decay $B^+ \to \overline{D}^{0} \ell^+ \nu_{\ell}$, with $\overline{D}^{0} \to K^+ \pi^-$, was used as a control mode. The reconstruction follows the same method except for adjusted mass requirements and adding a kaon. The measured branching fraction is $2.536 \pm 0.036 \pm 0.087 \%$. There is a $1.9 \sigma$ discrepancy with the world average~\cite{PDG} of $2.29 \pm 0.09 \%$ after adjusting to ${\cal B}(\Upsilon(4S) \to B^+B^-) = 0.513$. The measured branching fraction does however show good agreement with the previous measurement by Belle~\cite{BelleD}. The difference in central values of these measurements is $5\%$ which is used as the systematic uncertainty due to shape mismodeling in the fit variables and selection efficiency discrepancies; this is listed in Table~\ref{tab:syst:exp} as the control mode uncertainty.\\

\section{Results}
The branching fraction of $B^{+} \to \eta \ell^{+} \nu_{\ell}$ decay resulting from the fit is:
\begin{align}
    \textrm{by}\; \gamma\gamma     &: (2.91 \pm 0.64 \pm 0.32) \times 10^{-5},\\
    \textrm{by}\; \pi^+\pi^-\pi^0  &: (2.65 \pm 1.04 \pm 0.37) \times 10^{-5},
\end{align}
where the first uncertainty is the statistical uncertainty from the fit, while the second is systematic. Since the measurements of the branching fraction in the two $\eta$ decay modes are consistent with each other, we can average over both $\eta$ modes, assuming the statistical uncertainties to be uncorrelated and the systematic uncertainties to be fully correlated:
\begin{align}
  {\cal B}(B^{+} \rightarrow \eta \ell^{+} \nu_{\ell}) &= (2.83 \pm 0.55 \pm 0.34) \times 10^{-5}.
\end{align}
The branching fraction of $B^{+} \to \eta' \ell^{+} \nu_{\ell}$ decay resulting from the fit is:
\begin{align}
  {\cal B}(B^{+} \rightarrow \eta' \ell^{+} \nu_{\ell}) &= (2.79 \pm 1.29 \pm 0.30) \times 10^{-5}.
\end{align}
This result is compatible with and complements the earlier Belle result~\cite{BelleEta}, which uses hadronic tagging for the second $B$ meson in the event. Due to the different methods applied the statistical overlap between the two analyses is negligible and they can be considered independent. The branching fractions have been previously also measured by BaBar~\cite{BabarEtaLoose,BabarEtaUntagged,BabarEtaSemilep}, although Ref.~\cite{BabarEtaSemilep} reports a greater value. For the branching fraction for $B^{+} \to \eta' \ell^{+} \nu_{\ell}$ CLEO~\cite{CLEOEtaPrime} reports a value about an order of magnitude larger and incompatible with both this and the result from BaBar. The precision of this measurement is limited by the sample size. Significantly more precise results can therefore be expected in the future with the Belle II experiment at SuperKEKB.

\begin{table}[htb]
\centering
\caption{Breakdown of the systematic uncertainty in \%.}
\label{tab:syst:exp}
\begin{tabular*}{\linewidth}{@{\extracolsep{\fill}} @{\hspace{0.3cm}}lccc@{\hspace{0.3cm}}}
\hline\hline
Source & $\eta(\gamma\gamma)$ & $\eta(\pi^+\pi^-\pi^0)$ & $\eta^\prime$  \\ \hline
Statistical&22&39&46\\
Combined Systematic&11&14&11\\ \hline
\br{B^{\pm} \to X_{Bkg}}&2.4&1.7&1.3\\
\br{\eta^{(')} \to X}&0.51&1.2&1.7\\
$B \to D^{(*,**)} \ell^{+} \nu_{\ell}$ form factor&0.82&1.1&1.3\\ 
$B \to \eta^{(')} \ell^{+} \nu_{\ell}$ form factor&3.0&2.9&0.14\\
$B \to \omega \ell^{+} \nu_{\ell}$ form factor&0.81&2.1&2\\
$\overline{b} \to \overline{u} \ell^{+} \nu_{\ell}$ shape&0.39&0.15&0.21\\
Background with $K_L^0$&3.5&8.6&3.8\\
Continuum&0.2&0.62&0.63\\
$N_{B\overline{B}}$&1.4&1.4&1.4\\
\br{\Upsilon(4S) \to B^+B^-}&1.2&1.2&1.2\\
$\overline{b} \to \overline{u} \ell^{+} \nu_{\ell}$ yield&4.1&5.2&4.4\\
Monte Carlo statistics&0.86&1.3&2.3\\
Charged tracks&0.35&1.1&1.1\\
$\gamma$ detection&4.0&2.5&4.0\\
Electron PID&1.6&1.6&1.5\\
Muon PID&2.1&2.1&2\\
First $\pi^{\pm}$ PID&0&0.97&1.1\\
Second $\pi^{\pm}$ PID&0&1.3&2.2\\
Misidentified Leptons&4.3&5.5&2.3\\
Control Mode&5.0&5.0&5.0\\ \hline\hline
\end{tabular*}
\end{table}

\section{Acknowledgments}
We thank the KEKB group for the excellent operation of the
accelerator; the KEK cryogenics group for the efficient
operation of the solenoid; and the KEK computer group, and the Pacific Northwest National
Laboratory (PNNL) Environmental Molecular Sciences Laboratory (EMSL)
computing group for strong computing support; and the National
Institute of Informatics, and Science Information NETwork 5 (SINET5) for
valuable network support.  We acknowledge support from
the Ministry of Education, Culture, Sports, Science, and
Technology (MEXT) of Japan, the Japan Society for the 
Promotion of Science (JSPS), and the Tau-Lepton Physics 
Research Center of Nagoya University; 
the Australian Research Council including grants
DP180102629, 
DP170102389, 
DP170102204, 
DP150103061, 
FT130100303; 
Austrian Federal Ministry of Education, Science and Research (FWF) and
FWF Austrian Science Fund No.~P~31361-N36;
the National Natural Science Foundation of China under Contracts
No.~11435013,  
No.~11475187,  
No.~11521505,  
No.~11575017,  
No.~11675166,  
No.~11705209;  
Key Research Program of Frontier Sciences, Chinese Academy of Sciences (CAS), Grant No.~QYZDJ-SSW-SLH011; 
the  CAS Center for Excellence in Particle Physics (CCEPP); 
the Shanghai Pujiang Program under Grant No.~18PJ1401000;  
the Shanghai Science and Technology Committee (STCSM) under Grant No.~19ZR1403000; 
the Ministry of Education, Youth and Sports of the Czech
Republic under Contract No.~LTT17020;
Horizon 2020 ERC Advanced Grant No.~884719 and ERC Starting Grant No.~947006 ``InterLeptons'' (European Union);
the Carl Zeiss Foundation, the Deutsche Forschungsgemeinschaft, the
Excellence Cluster Universe, and the VolkswagenStiftung;
the Department of Atomic Energy (Project Identification No. RTI 4002) and the Department of Science and Technology of India; 
the Istituto Nazionale di Fisica Nucleare of Italy; 
National Research Foundation (NRF) of Korea Grant
Nos.~2016R1\-D1A1B\-01010135, 2016R1\-D1A1B\-02012900, 2018R1\-A2B\-3003643,
2018R1\-A6A1A\-06024970, 2018R1\-D1A1B\-07047294, 2019K1\-A3A7A\-09033840,
2019R1\-I1A3A\-01058933;
Radiation Science Research Institute, Foreign Large-size Research Facility Application Supporting project, the Global Science Experimental Data Hub Center of the Korea Institute of Science and Technology Information and KREONET/GLORIAD;
the Polish Ministry of Science and Higher Education and 
the National Science Center;
the Ministry of Science and Higher Education of the Russian Federation, Agreement 14.W03.31.0026, 
and the HSE University Basic Research Program, Moscow; 
University of Tabuk research grants
S-1440-0321, S-0256-1438, and S-0280-1439 (Saudi Arabia);
the Slovenian Research Agency Grant Nos. J1-9124 and P1-0135;
Ikerbasque, Basque Foundation for Science, Spain;
the Swiss National Science Foundation; 
the Ministry of Education and the Ministry of Science and Technology of Taiwan;
and the United States Department of Energy and the National Science Foundation.

\appendix*
\section{BDT training variable distributions}

\begin{figure*}[htpb]
\centering
\subfloat{\includegraphics[width=0.64 \textwidth]{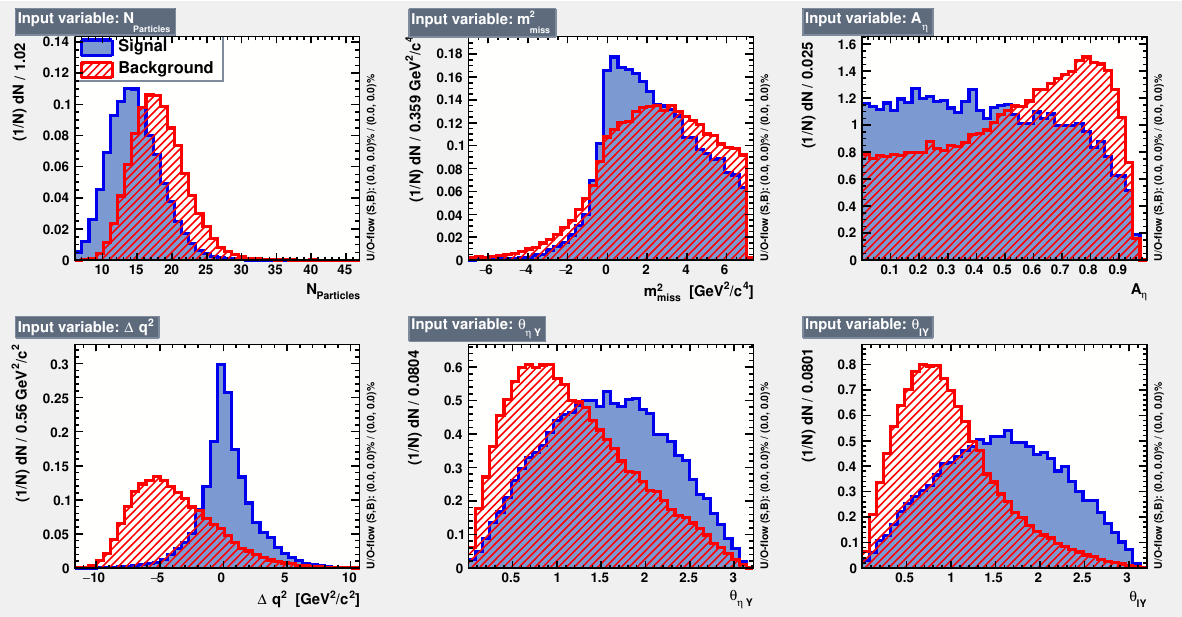}}\\
\subfloat{\includegraphics[width=0.64 \textwidth]{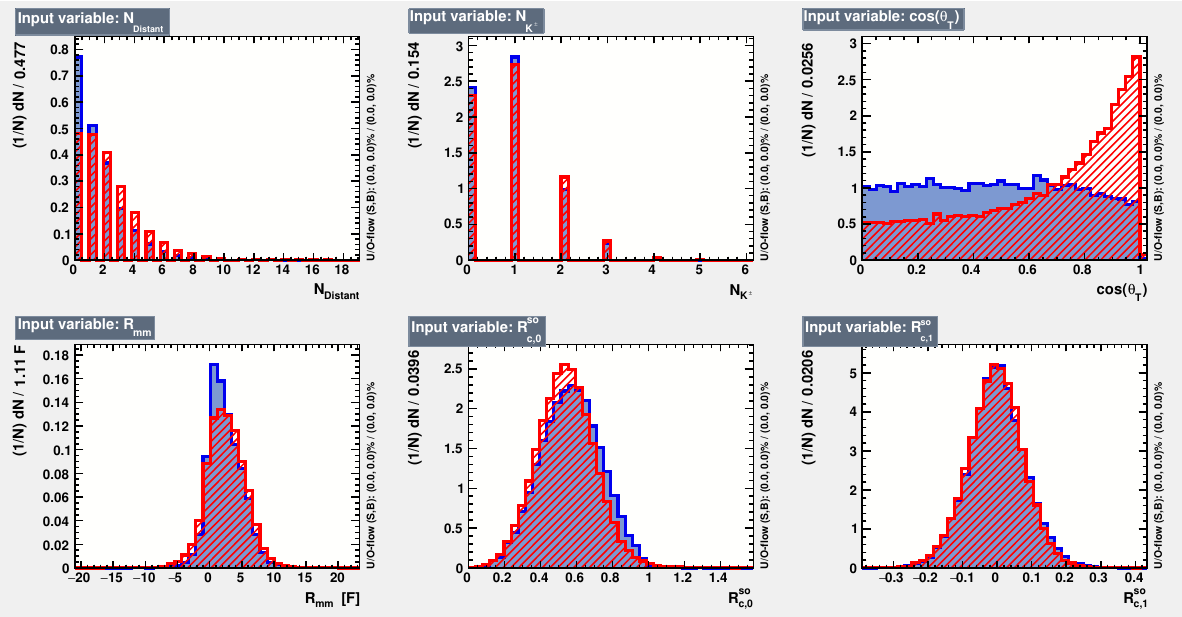}}\\
\subfloat{\includegraphics[width=0.64 \textwidth]{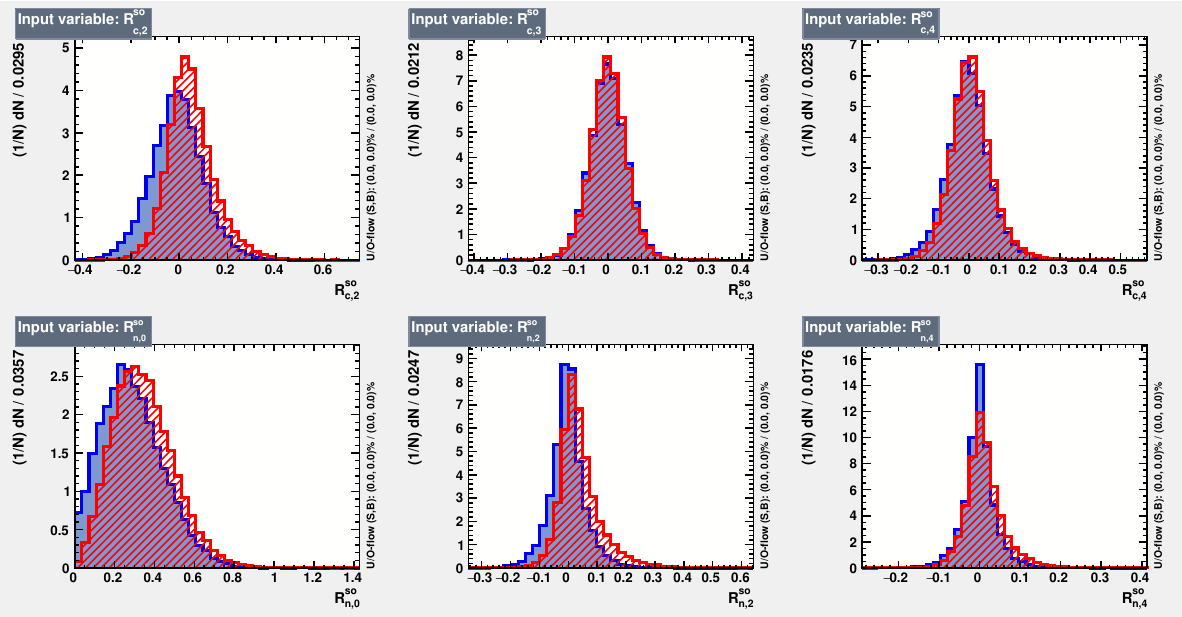}}\\
\subfloat{\includegraphics[width=0.64 \textwidth]{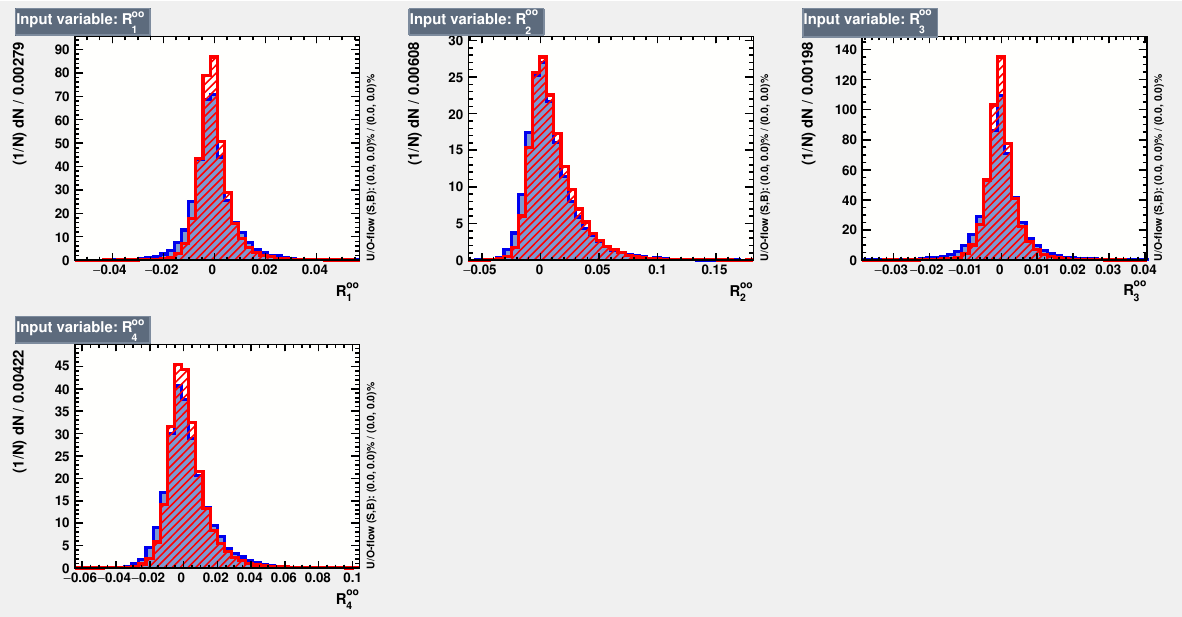}}
\caption{BDT training variables in the channel $\eta\to\gamma\gamma$ }
\label{bdtvars1}
\end{figure*}

\begin{figure*}[htpb]
\centering
\subfloat{\includegraphics[width=0.64 \textwidth]{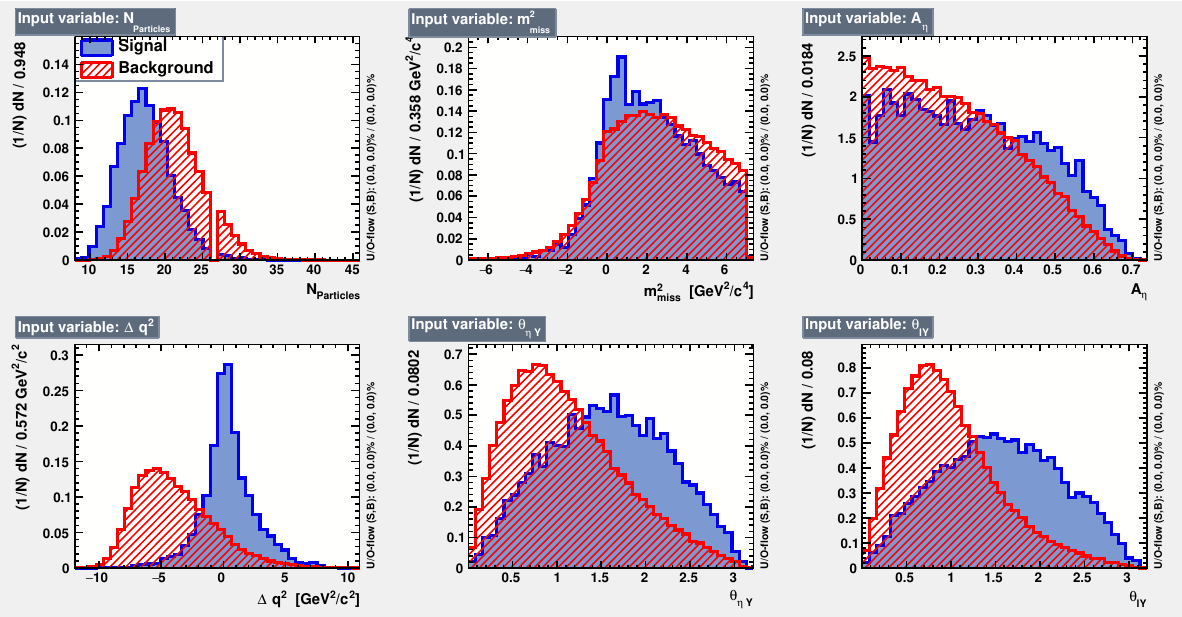}}\\
\subfloat{\includegraphics[width=0.64 \textwidth]{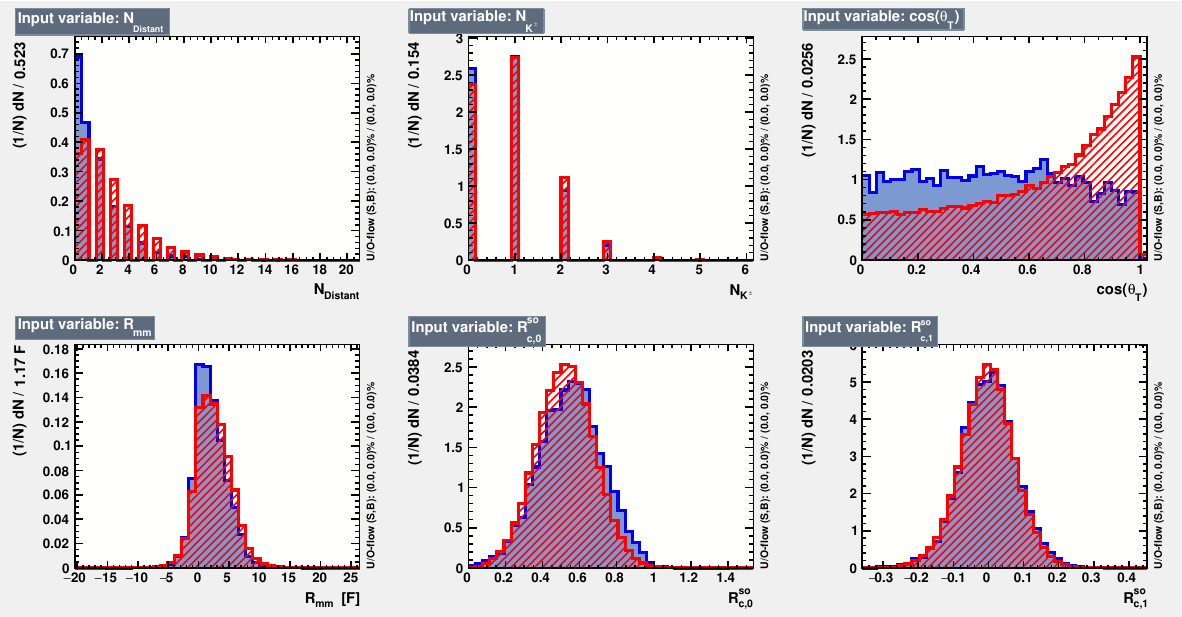}}\\
\subfloat{\includegraphics[width=0.64 \textwidth]{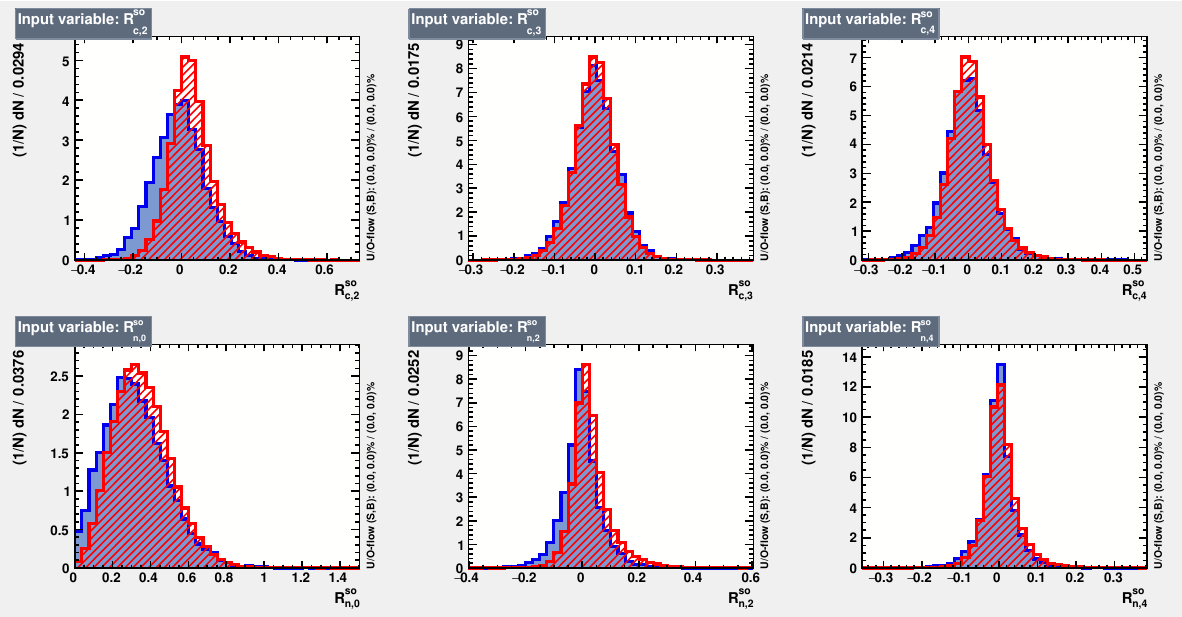}}\\
\subfloat{\includegraphics[width=0.64 \textwidth]{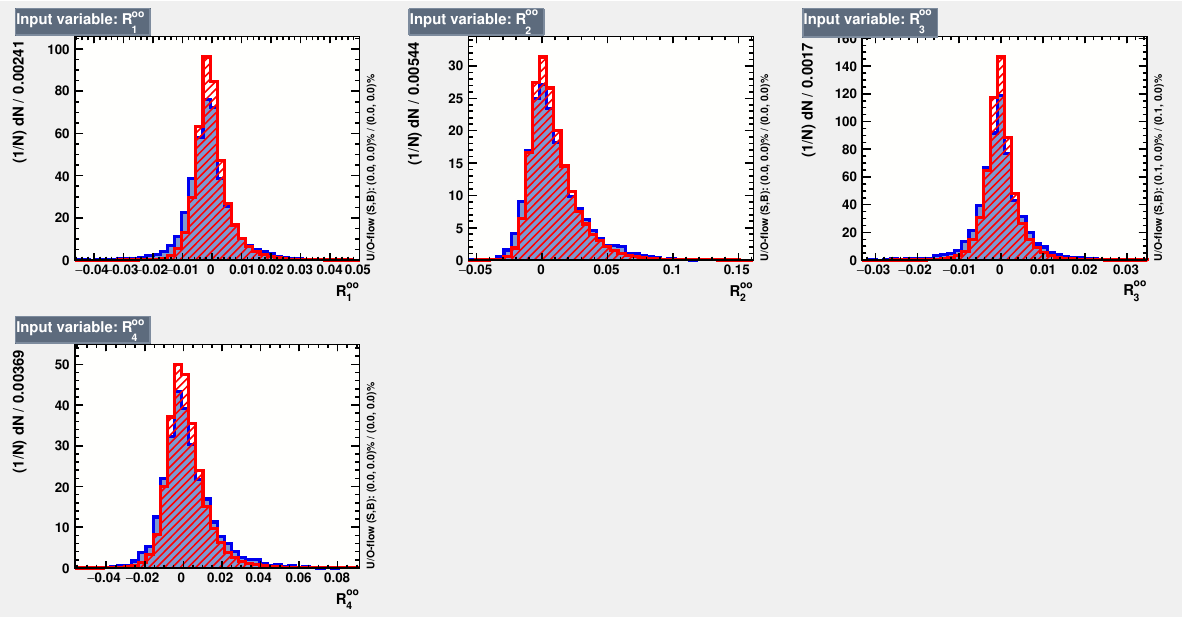}}
\caption{BDT training variables in the channel $\eta\to\pi^+\pi^-\pi^0$ }
\label{bdtvars2}
\end{figure*}

\begin{figure*}[htpb]
\centering
\subfloat{\includegraphics[width=0.64 \textwidth]{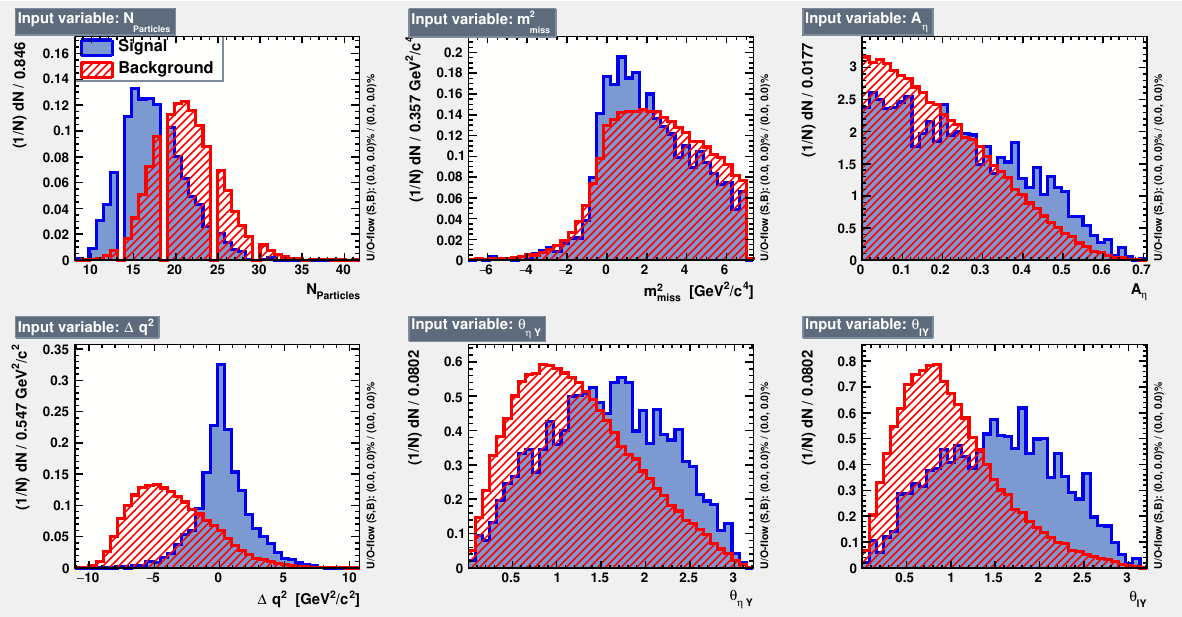}}\\
\subfloat{\includegraphics[width=0.64 \textwidth]{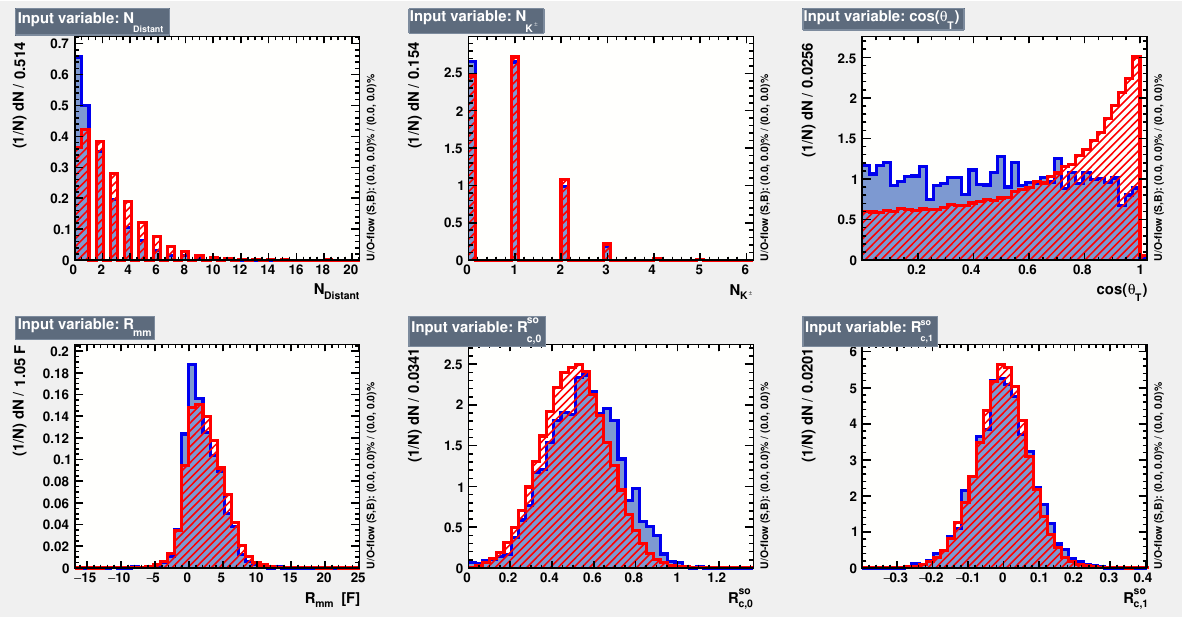}}\\
\subfloat{\includegraphics[width=0.64 \textwidth]{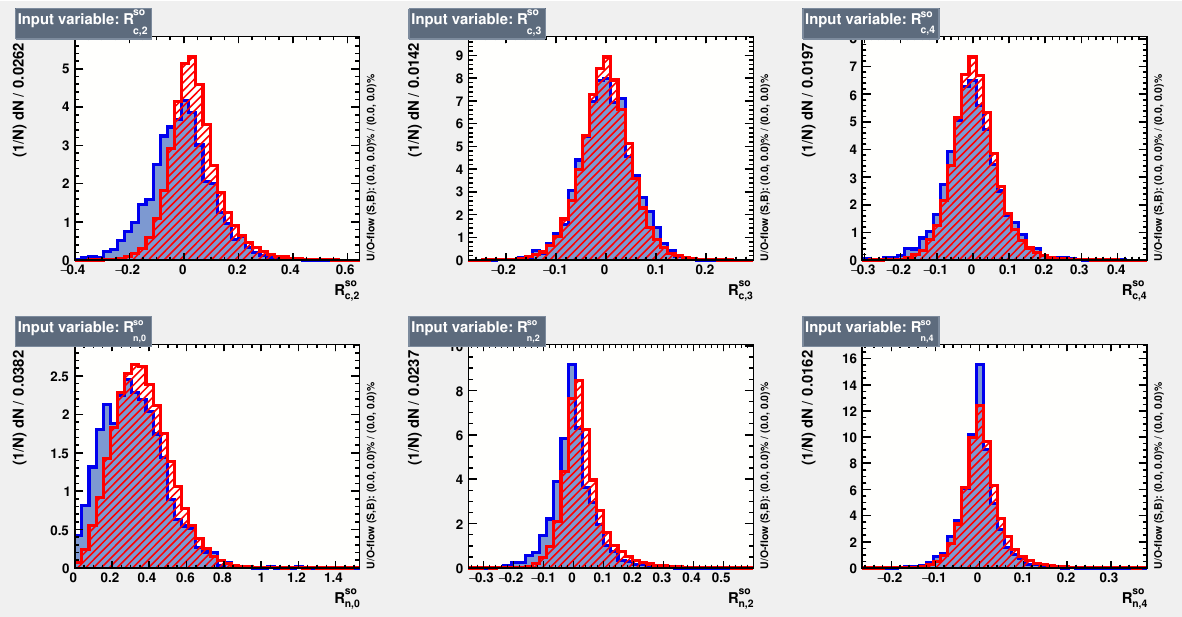}}\\
\subfloat{\includegraphics[width=0.64 \textwidth]{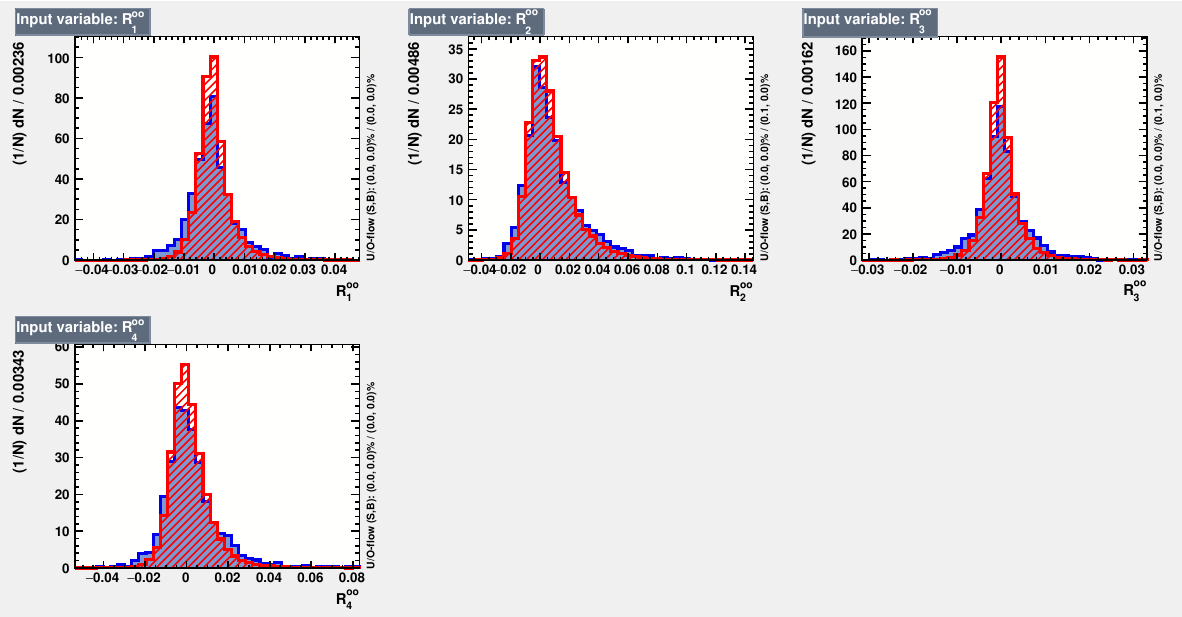}}
\caption{BDT training variables in the channel $\eta^\prime\to\pi^+\pi^-\eta(\gamma\gamma)$ }
\label{bdtvars6}
\end{figure*}


\begin{thebibliography}{99}
\bibitem{CC}
Throughout this paper, 
the inclusion of the charge-conjugate decay mode is implied
unless otherwise stated.
\end{thebibliography}
\end{document}